\documentclass[aps,twocolumn,showpacs,superscriptaddress,pre,floatfix]{revtex4-2}
\usepackage{amsmath}
\usepackage{amssymb}
\usepackage{epsfig}
\usepackage{multirow}
\usepackage{graphicx}
\usepackage{tikz}
\usepackage{filemod}
\usepackage{float}
\usepackage{textcomp}
\usepackage{datetime}
\usepackage{mathtools}
\usepackage{hyperref}
\hypersetup{
    colorlinks,
    linkcolor={blue!80!black},
    citecolor={blue!80!black},
    urlcolor={blue!80!black}
}

\usepackage{stackengine}

\graphicspath{{./figures/}}

\def\SV{\operatorname{SV}}
\def\sv{\operatorname{sv}}

\usepackage{enumitem}
\usepackage{makecell}
\usepackage{pdfcomment}

\begin{document}

\title{Resampling schemes in population annealing: Numerical and theoretical results}
\date{\today}

\author{Denis Gessert}
\email{denis.gessert@itp.uni-leipzig.de}
\affiliation{Centre for Fluid and Complex Systems, Coventry University, Coventry CV1~5FB, United Kingdom}
\affiliation{Institut f\"{u}r Theoretische Physik, Leipzig University, IPF 231101, 04081 Leipzig, 
  Germany}

\author{Wolfhard Janke}
\email{wolfhard.janke@itp.uni-leipzig.de}
\affiliation{Institut f\"{u}r Theoretische Physik, Leipzig University, IPF 231101, 04081 Leipzig, 
  Germany}

\author{Martin Weigel}
\email{martin.weigel@physik.tu-chemnitz.de}
\affiliation{Institut f\"ur Physik, Technische Universit\"at Chemnitz, 09107 Chemnitz, Germany}

\begin{abstract}
    The population annealing algorithm is a population-based equilibrium version of simulated annealing. It can sample thermodynamic systems with rough free-energy landscapes more efficiently than standard Markov chain Monte Carlo alone. A number of parameters can be fine-tuned to improve the performance of the population annealing algorithm. While there is some numerical and theoretical work on most of these parameters, there appears to be a gap in the literature concerning the role of resampling in population annealing which this work attempts to close.
    The two-dimensional Ising model is used as a benchmarking system for this study. At first various resampling methods are implemented and numerically compared. In a second part the exact solution of the Ising model is utilized to create an artificial population annealing setting with effectively infinite Monte Carlo updates at each temperature. This limit is first performed on finite population sizes and subsequently extended to infinite populations. This allows us to look at resampling isolated from other parameters. Many results are expected to generalize to other systems.
\end{abstract}


\maketitle



\section{Introduction}

Without doubt the enormous increase in computing power over the past decades has paved the way to tackle ever more challenging problems. 
In parallel to this explosion in raw computing power, the design of efficient algorithms and their further refinement have turned out to be crucial for solving many hard computational tasks. One class of such problems requires the simulation of complex systems with rugged free-energy landscapes, such as spin glasses, polymers and frustrated systems~\cite{Janke2007}. In the recent past it has been shown that the population annealing (PA) framework can be rather successful in treating the aforementioned systems~\cite{Hukushima2003,Machta2010,Amey2018,Christiansen2019a}.

PA is a simulation framework in which a population of $R$ replicas, i.e., $R$ configurations of the model under study, is collectively cooled from an initial high temperature to a final low temperature. Replicas evolve independently except at each temperature step where the population is reweighted and resampled (see Sec.~\ref{sec:algoPA} for details). Typically Markov chain Monte Carlo (MCMC) is used to evolve replicas between temperature steps but any update (including molecular dynamics) which is suitable for the studied model may be used~\cite{Amey2021,Christiansen2019a}. Each time the temperature is lowered, replicas acquire an importance weight resulting from the change in temperature. 
Resampling, the subject of this study, in essence then is the ``translation'' of real-valued weights to integer numbers of copies to be made of each replica as the temperature is lowered to again evenly distribute the weight among all members of the population.

Over the past decade and a half the increase in computing power has no longer translated into an improvement in single core performance but rather a constant increase of the number of computing cores available. PA is almost trivially parallel and has no theoretical limitations on the level of parallelism~\cite{Weigel2017,Shchur2018} which means it is well suited to run on modern computing hardware that is becoming more and more parallel. Parallelizability is perhaps the unique selling point for PA over related approaches with otherwise similar performance~\cite{weigel:18}. For instance, a somewhat similar algorithm is parallel tempering~\cite{Hukushima1996} for which, however, the potential degree of parallelism is rather limited~\cite{Bittner2008}.

Note that without weights and resampling PA essentially reduces to performing $R$ independent simulated annealing (SA) runs~\cite{Weigel2021}. Keeping track of the weights is necessary if correctly weighted thermal averages are to be taken over the population (which is normally not the focus of attention in simulated annealing)~\cite{Kirkpatrick1983}. However, if weights are never rebalanced down to low temperatures, then few replicas will carry most of the weight, such that most computational resources are spent on replicas that do not contribute to measurements~\cite{Gordon1993}. Thus, resampling contributes toward the ongoing task of equilibration and importance sampling of the system. 

Although it is well understood that resampling is a crucial part in the PA framework~\cite{Weigel2021} and despite the fact that numerous different resampling schemes have been used in PA in the past~\cite{Hukushima2003,Machta2010,Barash2017}, to the best of our knowledge, the effect of the chosen resampling method on the quality of the data obtained in PA has not yet been studied systematically. In the present work we attempt to fill this gap.
Besides providing guidance regarding the question which method is preferable, we quantify the noise that enters simulations through resampling and thus identify scenarios in which the chosen resampling method matters most, thus revealing a tight connection between resampling and the temperature schedule.

The rest of this paper is organized as follows. In Sec.~\ref{sec:algo} we describe the PA algorithm as well as the different resampling methods we consider. Section~\ref{sec:modelSimulationObservables} contains an outline of the simulation details, in particular the quantities we use to compare the different methods. Our results are split into two parts, presented in Secs.~\ref{sec:firstNumericalObservations} and \ref{sec:thetaInfLimit}. 
In the former we discuss numerical results from PA simulations using various resampling methods and otherwise constant parameters, while in the latter we study the effect of the chosen resampling method in well-equilibrated systems using the perfectly equilibrated Ising model as an artificial example. Finally, Sec.~\ref{sec:conclusion} contains our conclusions.


\section{Algorithm}
\label{sec:algo}
\subsection{Population annealing}
\label{sec:algoPA}
As already mentioned above, PA is an algorithmic framework in which a population of replicas is sequentially cooled. Each temperature step is followed by a population control move. Finally, to additionally equilibrate the system, one also performs a number of single-replica MCMC moves at each temperature. Through the temperature steps each replica $k=1,\ldots,R_i$ at temperature step $i$ acquires a weight $W_k^{(i)}$, and these weights are rebalanced through resampling. The algorithm can be summarized as follows:

\begin{enumerate}
    \item Initialize the population of $R$ replicas at the starting inverse temperature $\beta_0 = 0$. In some cases nonzero inverse temperature has to be chosen~\cite{Christiansen2019a}. Set iteration counter $i\leftarrow 0$. Set all weights ${W^{(i)}_k \leftarrow 1}$.
    \item Make an inverse temperature step $\beta_i \to \beta_{i+1}$ unless the stopping inverse temperature $\beta_s$ is reached, i.e., unless $\beta_i \geq \beta_s$ : \begin{enumerate}[label=(\alph*),ref=\theenumi{}\,(\alph*)]
        \item \label{item:pa_calcWeights} Calculate the modification of the (unnormalized) Boltzmann weights $W_k^{(i)}$ for each replica $k$ with energy $E_k$ through the temperature change, i.e., 
        \begin{equation}
            W_k^{(i+1)} = W_k^{(i)} e^{-\Delta\beta E_k}\,,
        \end{equation}
        where $\Delta\beta\coloneqq\beta_{i+1}-\beta_i$ is the inverse temperature step.
        \item \label{item:pa_resample} Resample the population according to the computed weights $W_k^{(i+1)}$, that is make on average 
        \begin{equation}
            \left.\tau_k = R W_k^{(i+1)} \middle/ \sum_{j=1}^{R_i} W_j^{(i+1)}\right.
        \end{equation}
        copies of replica~$k$, where $R_i$ is the population size at $\beta_i$. Set all weights ${W^{(i+1)}_k \leftarrow 1}$.
        \item Increment the iteration counter, $i \leftarrow i+1$.
        \item Perform $\theta$ MCMC sweeps on each replica.
        \item Calculate estimates for observables through population averages, i.e.,
        \begin{equation}
          \hat{\mathcal{O}}^{(i)} = \sum_{k=1}^{R_i} \mathcal{O}_k^{(i)} W_k^{(i)} / \sum_{k=1}^{R_i} W_k^{(i)}\,.
        \end{equation}
        \item Go to 2.
    \end{enumerate}
\end{enumerate}

The three major parameters that can be adjusted to optimize PA performance are the (target) population size~$R$, the number of updates in the equilibration routine~$\theta$ and the inverse temperature step~$\Delta\beta$. Some guidelines for their choice have previously been discussed by some of us~\cite{Weigel2021}.
In essence, one should choose $\theta$ large enough to ascertain a sufficient degree of equilibration if an efficient MCMC algorithm is available and put the remaining computing resources into choosing a population size as large as easily feasible, typically of the order of at least a few thousand replicas.
The annealing schedule $\{\beta_i\}$ is recommended to be chosen adaptively~\cite{Barash2017, Christiansen2019, Weigel2021}.

Note, that when the weights $W_k^{(i)}$ are reset in every iteration (as is the case here), then they can be absorbed into the expression for $\tau_k$, i.e.,
\begin{equation}
    \left.\tau_k = R e^{-\Delta \beta E_k} \middle/ \sum_{j=1}^{R_i} e^{-\Delta \beta E_j}\,.\right. \label{equ:tau}
\end{equation}
In this case, population averages can simply be calculated as $\hat{\mathcal{O}}^{(i)} = \sum_{k=1}^{R_i} \mathcal{O}_k^{(i)} / R_i$.
More generally, instead of resetting weights to unity, one can resample such that weights after the resampling are set following a rule $W_k \leftarrow g(W_k)$ for some $g(x)$, e.g., $g(x) = \sqrt{x}$ (see ch.\,11.3.1 in Ref.~\cite{Doucet2001}). Such choices amounting to a trade-off between importance sampling and increased correlations in the population will not be considered further in the present paper, however.

\subsection{Resampling methods}
\label{sec:resamplingMethods}

We now turn specifically to the resampling procedure of step~\ref{item:pa_resample} in the algorithm. It is a random process that is geared toward removing the imbalance among the replica weights computed in step \ref{item:pa_calcWeights} which is a consequence of the variation in configurational energy. Equation~\eqref{equ:tau} only determines the expected number of copies, i.e., if $r_k^{(i)}$ is the number of copies made of replica $k$ at temperature step $i$, then we demand that
\begin{equation}
\langle r_k^{(i)}\rangle = \tau_k^{(i)}\,.
\end{equation}
One hence has a wide freedom in choosing the distribution of $r_1^{(i)},\ldots,r_{R_i}^{(i)}$ as only its first moment is fixed.
In total, the $R_i$ configurations are resampled into $R_{i+1} = \sum_{k=1}^{R_i} r_k^{(i)}$ replicas~\footnote{Furthermore, instead of restricting the pool of configurations from which to resample to the ones obtained after all equilibration steps, one can extend the pool by the configurations sampled during the equilibration~\cite{Rose2019,Amey2021}.} such that on average each replica~$k$ is copied $\tau_k$ many times; as a result, $R_{i+1}$ may differ from the target population size $R$. In fact, we will distinguish between methods that preserve a constant population size throughout and those which have a fluctuating population size. On distributed architectures the former may be desirable as they allow to guarantee that every compute node has the same number of replicas.
The (resampled) $R_{i+1}$ configurations then carry equal weights $W_k^{(i+1)} = 1$.
In the following we will first discuss the methods with fixed and then the ones with fluctuating population size.
Note that some of the methods with constant population size we discuss here have been studied previously in the context of particle filtering~\cite{Cappe2005}. 

\begin{figure*}
    



    
    
    
    \includegraphics{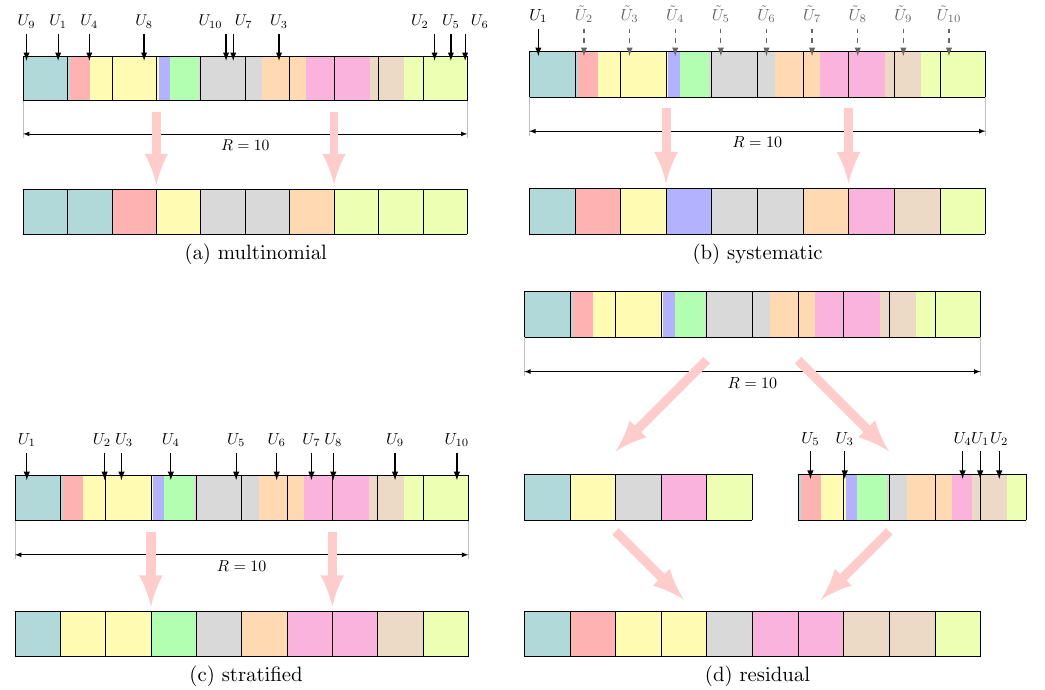}

    \caption{Visualization of various population-size preserving resampling methods. Colored boxes correspond to different replicas $k$ and their lengths are proportional to $\tau_k$. The number of copies of replica~$i$ is determined by the number of arrows in (colored) box $k$. The resampled population is shown at the bottom for each method. The algorithms for each method are explained in the main text.}
    \label{fig:VisResamplingPreservedPopulation}
\end{figure*}

For \emph{fixed population size} the distribution $P(r^{(i)}_1,\ldots,r^{(i)}_{R_i})$ does not factorize and hence the set $\{r_k\}$ has to be drawn at once. For a full mathematical description of such methods we refer to ch. 4.3.1 of Ref.~\cite{Doucet2001} as well as Ref.~\cite{doucet2009tutorial}, and we will here only provide an algorithmic explanation of the following approaches, 
\begin{enumerate}[label=(\alph*)]
    \item multinomial resampling,
    \item systematic resampling,
    \item stratified resampling, and
    \item residual resampling.
\end{enumerate}
To the best of our knowledge, this is the first time that population-size preserving methods other than multinomial resampling are studied in the context of PA.
Previous work~\cite{Hukushima2003,Christiansen2019a,Machta2011} always relied on multinomial resampling when requiring a constant population size. 

The methods can best be explained through the geometric illustration provided in Fig.~\ref{fig:VisResamplingPreservedPopulation}. The desired number of copies $\tau_k$ for each replica $k$ is visualized by horizontal stacked bar charts. By design the $\tau_k$'s add up to $R$, as is indicated by the double-arrow line of length $R$ and the grid of unit squares. Each method (with the exception of residual resampling) can then be explained by a protocol of placing $R$ arrows at random on the interval $[0,R]$. The number of arrows falling onto a colored box with length $\tau_k$ then determines the number of copies made, $r_k$. Finally, he bottom bar shows the population after resampling.

In \emph{multinomial resampling} the position $U_i$ of each of the $R$ arrows is chosen uniformly at random from the interval $[0,R]$, see Fig.~\ref{fig:VisResamplingPreservedPopulation}(a). Hence, the $U_i$'s appear to be out of order as compared to systematic and stratified resampling. Formally, this corresponds to drawing from a multinomial distribution with $R$ trials, $R$ mutually exclusive events and event probabilities $p_1, \dots, p_{R}$ equal to $\tau_1 / R, \dots, \tau_R / R$ as in Eq.~(\ref{equ:tau}).
%
%
In \emph{systematic resampling} [see Fig.~\ref{fig:VisResamplingPreservedPopulation}(b)] the position of the first arrow $U_1$ is drawn uniformly at random from the interval $[0,1]$ (first square). The positions $\tilde U_2, \dots, \tilde U_R$ of the remaining $R-1$ arrows are given by $\tilde U_k = U_1 + k -1$. Remarkably, this method only uses one random number for the resampling of the population, as is illustrated by the use of $\tilde U_k$ instead of $U_k$ and the dashed arrows as well as the lighter color in Fig.~\ref{fig:VisResamplingPreservedPopulation}(b).
Similarly, in \emph{stratified resampling} [see Fig.~\ref{fig:VisResamplingPreservedPopulation}(c)] only one arrow is placed per square.
However, here the arrows are not spaced equidistantly but rather placed with uniform probability on each square, i.e., $U_i \sim \mathcal U ([i-1,i])$.
Last, in \emph{residual resampling} [see Fig.~\ref{fig:VisResamplingPreservedPopulation}(d)] at first each replica is copied $\lfloor\tau_k\rfloor$ times, where $\lfloor x \rfloor$ denotes the largest integer smaller than or equal to $x$, i.e., rounding down.
The population is brought to its original size by multinomially drawing from the residuals, that is by performing multinomial resampling where $\tau_k$ is replaced by $\tau_k - \lfloor\tau_k\rfloor$. Note that the sum of the residuals is a random variable and can take any nonnegative integer value up to $R-1$. Hence, residual resampling uses less random numbers than multinomial and stratified and the actual number is a random variable.
In fact, instead of multinomially sampling the residuals one may choose to use the systematic or stratified resampling method instead. This is then commonly referred to as systematic residual, respectively, stratified residual resampling.
Here, we only discuss (multinomial) residual resampling but extensions to different resampling methods are straightforward.

Clearly, all methods shown in Fig.~\ref{fig:VisResamplingPreservedPopulation} distribute $R$ arrows and thus the resampled population is guaranteed to have the target size. Additionally, in all methods the expected numbers of arrows in each colored box (i.e., per replica) is proportional to its size, i.e., $\tau_k$, and hence the original constraint of $\langle r_k \rangle = \tau_k$ is also satisfied.

When allowing a \emph{fluctuating population size} one can use factorized distributions for $P(r^{(i)}_1,\ldots,r^{(i)}_{R_i})$, in which the number of copies $r_k$ of one replica~$k$ is chosen only based on $\tau_k$ of the same replica~$k$, i.e.,
\begin{equation}
    P(r^{(i)}_1,\ldots,r^{(i)}_{R_i}) = \prod_{k=1}^{R_i} P_{\tau^{(i)}_k}(r^{(i)}_k)\,.
\end{equation} In this case, resampling can easily be implemented in parallel~\cite{Barash2017} as the resampling method is completely described by a univariate distribution $P_{\tau_k}(r_k = j)$. As a consequence, the new population size $R_{i+1}$ (i.e., the sum of all $r_k$'s) itself becomes a random variable. Hence, the population size fluctuates with time. In this group we consider
\begin{itemize}
    \item[(e)] nearest-integer resampling~\cite{Wang2015} and 
    \item[(f)] Poisson resampling~\cite{Machta2010}
\end{itemize} 
given by 
\begin{equation}
    P_{\tau_k}(r_k = j) = \begin{cases}
        \tau_k - \lfloor\tau_k \rfloor & \text{if } j = \lfloor\tau_k \rfloor + 1 \\
        1 - (\tau_k - \lfloor\tau_k \rfloor) & \text{if } j = \lfloor\tau_k \rfloor \\
        0 & \text{else}
    \end{cases}
\end{equation}
for nearest integer and
\begin{equation}
    P_{\tau_k}(r_k = j) = \frac{\tau_k^{j}}{j!} \, e^{-\tau_k }\hspace{2.1cm}
\end{equation}
for Poisson, respectively.

Besides the question of whether a method preserves population size, the most notable difference among the above approaches is by how much $r_k$ can differ from $\tau_k$ or, more quantitatively, what the variance of $r_k$ is. As is shown in Appendix~\ref{app:samplingVarianceCalc}, it is possible (under mild assumptions) to exactly calculate this variance as a function of $\tau_k$. This quantity will play a crucial role in Sec.~\ref{sec:thetaInfLimit}. Apart from the distributions (a)--(f) above, any probability distribution with nonnegative integer support and adjustable mean (see, for example, Ref.~\cite{DeGroot2012}), such as the geometric or the Pascal distribution, would also result in a valid resampling method, but these lead to even larger sampling variances and hence are of no practical relevance.

\section{Model, simulation details, and observables}
\label{sec:modelSimulationObservables}
\subsection{Ising model}
In this work we consider the Ising model in the absence of an external magnetic field, corresponding to the Hamiltonian
\begin{equation}
    \mathcal{H} = -J \sum_{\langle ij \rangle} \sigma_i \sigma_j\label{eq:Hamiltonian}\,,
\end{equation}
where $\sigma_i\in\{-1,1\}$ are the spin variables and the sum is over nearest-neighbor interactions only. We choose $J=1$ and $k_B = 1$ to fix units and use a $L \times L$ square lattice with periodic boundary conditions.

Owing to the availability of exact results~\cite{Onsager1944,Kaufmann1949,Beale1996}, this model has become a standard benchmark system for PA~\cite{Barash2017,Weigel2021,Ebert2022} and many other algorithms. In particular, this allows for an easy way to differentiate systematic from statistical errors. An additional advantage of the two-dimensional Ising model is the availability of exact results for finite systems~\cite{Kaufmann1949,Beale1996}, particularly the exact energy density of states~\cite{Beale1996}, which allows us to separate the effect of resampling from MCMC as we will explain in Sec.~\ref{sec:thetaInfLimit}. 
\subsection{Simulation details}
\label{sec:simulationDetail}
We use the publicly available code from Ref.~\cite{Barash2017} to obtain the numerical data presented in Sec.~\ref{sec:firstNumericalObservations}. This implementation runs on a single GPU and is highly parallel. All calculations of weights, resampling~\footnote{In the implementation of the resampling methods with constant population size some parts of the resampling step are performed sequentially.} and measuring observables are done in parallel,
with one thread per replica. Spin-flip updates, on which typically most wall-clock time is spent, are further parallelized to a sub-replica level by employing a checkerboard domain-decomposition. When global summation is
needed, such as for the normalization of weights, this is done efficiently by first calculating partial sums of each thread block on subsets of the total population and then summing over the partial sums by using atomic operations provided by the CUDA toolkit. Necessary extensions of the scheme described in Ref.~\cite{Barash2017}, such as the calculations required for the realization of the different resampling techniques discussed in Sec.~\ref{sec:resamplingMethods}, were implemented in the same spirit.





The question we set out to answer is which resampling method has the best PA performance. In principle, one may expect this question to be decided in a trade-off between run time and accuracy. In most practical scenarios, however, run time in PA is vastly dominated by Monte Carlo (MC) moves such that even for the Ising model where spin updates are quick, resampling often takes less than 1\% of the overall run time (cf.\ the inset of Fig.~19 in Ref.~\cite{Weigel2021}). Thus, we focus on observables which measure the quality of the data obtained through PA and will not compare run times.

\subsection{Observables}

The most immediately suitable quantities for judging the performance of PA are the systematic and statistical errors. We looked at various moments of the energy and magnetization distributions and present here the errors in estimating the specific heat, since the exact solution~\cite{Kaufmann1949} allows us to readily evaluate the systematic error and since this is where we found the strongest differences among the resampling methods we studied.

Another useful quantity to compare different resampling methods is the average quadratic deviation between the expected and the actually generated number of copies, namely the \emph{sampling variance}~\cite{Li2015a}, i.e.,
\begin{equation}
    \SV = \frac {1 }{R_i} \sum_{k=1}^{R_i}(r_k  - \tau_k)^2\,.
\end{equation}
This quantity is of particular interest as it is a direct measure of how much additional noise enters the PA simulation through the resampling step.

As a consequence of the resampling step replicas descending from the same replica at a previous temperature will in general be correlated. This is illustrated by the notion of \emph{families}~\cite{Wang2015}: Two replicas $a$ and $b$ are said to belong to the same family if they both are descendants of the same initial replica $k$ (at $\beta_0$). In contrast, two replicas from \emph{different} families are guaranteed to be uncorrelated.

The authors of Ref.~\cite{Wang2015} quantify the family distribution by the \emph{replica-averaged family size }$\rho_t$, 
    \begin{equation}
        \rho_t = R \sum_{k=1}^{R} \mathfrak{n}_k^2\,, \label{equ:rho_t}
    \end{equation}
the \emph{entropic family size},
    \begin{equation}
        \rho_s = R \exp\left( \sum_{k=1}^R \mathfrak{n}_k \ln \mathfrak{n}_k\right)\,,
    \end{equation}
and the \emph{number of families},
    \begin{equation}
        f = \sum_{k=1}^{R} \min\{1,R\mathfrak{n}_k \}\,,
    \end{equation}
where $\mathfrak{n}_k$ is the fraction of replicas descending from the initial replica $k$. Note that while $k$ runs from 1 to $R$, most $\mathfrak{n}_k$ are zero. As is shown in Ref.~\cite{Wang2015}, these quantities are closely related to each other. For reasons outlined in Appendix~\ref{app:RhoTName} we use a different name for $\rho_t$ than previous authors.

Since, by construction, replicas from different families are uncorrelated, the above family quantities provide an upper bound for the population's correlation and thus an upper bound for error bars. This, of course, neglects the decorrelating effect of MCMC updates and hence largely overestimates the actual correlation~\cite{Weigel2021}. Most notably we can see this in the artificial PA setting described in Sec.~\ref{sec:thetaInfLimit}, as there by design all replicas are uncorrelated and yet the family quantities at the end of each simulation would suggest strong correlation, i.e., $\rho_t$, $\rho_s$, $R/f\gg 1$.

Recently, Weigel \emph{et al.}~\cite{Weigel2021} addressed this issue and developed a method to quantify PA correlations that takes into account the effect of MCMC updates. They observed that correlations in a PA population are similar in nature to the ones in MCMC time series, provided that copies of one replica are placed adjacently in the resampled population. This is demonstrated by the overlap correlation $C(i,j)$ of two replicas $i$ and $j$. They show that $C(i,j)$ in a large population only depends on the distance $|i-j|$, and it decays exponentially for large enough distance $|i-j|$, i.e., $C(i,j) = C(|i-j|) \propto \exp(-|i-j|/\tau_\text{exp})$ --- analogous to the well-known two-time correlation $C(t_1,t_2) \propto \exp(-|t_1-t_2|/\tilde\tau_\text{exp})$ in MCMC. 

The authors of Ref.~\cite{Weigel2021} further use a binning analysis of observables such as energy and magnetization in replica space to obtain error bars and show that they are compatible with error bars calculated from independent PA simulations.
From this, a measure of effective population size $R_\text{eff}$ is defined through
\begin{equation}
    R_\text{eff}(\mathcal O) = \frac{\sigma^2(\mathcal{O})}{\sigma^2_R(\bar{\mathcal{O}})}\,,
\end{equation}
where $\sigma^2(\mathcal{O})$ is the variance of the observable~$\mathcal O$ and $\sigma^2_R(\bar{\mathcal{O}})$ is the variance of its mean. The first is straightforward to calculate and the latter can be obtained through (jackknife) binning~\cite{Efron1982}, i.e.,
\begin{equation}
    \hat \sigma^2_R(\bar{\mathcal O}) = \frac{1}{n (n-1)} \sum_{i=1}^n \left(\mathcal{O}_i^{(n)} - \bar{\mathcal O}\right)^2\,,
\end{equation}
where $n$ blocks are chosen large enough that bins can be assumed to be uncorrelated and $\mathcal{O}_i^{(n)}$ is the mean of the $i$th bin. We will use $R_\text{eff}(E)$ and $R_\text{eff}(M)$ as further means to benchmark different resampling methods.

\begin{figure}
    \centering
    \includegraphics[width=0.9\columnwidth]{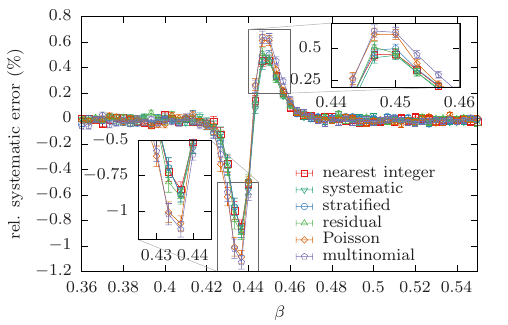}
    \includegraphics[width=0.9\columnwidth]{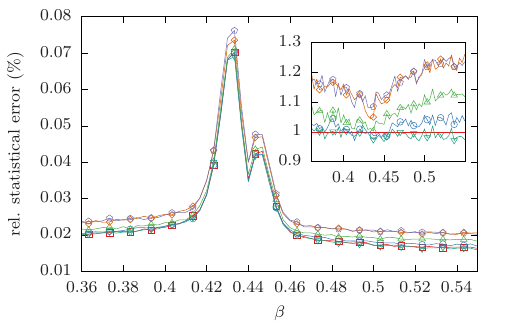}
    \caption{Relative systematic (upper panel) and statistical (lower panel) error of the specific heat measured after the resampling step for various resampling methods. The data are extracted from PA simulations with $R=20\,000$, $\theta=5$, and $\beta_i = i / 300$. Upper panel: Systematic errors hardly differ between different resampling schemes. Lower panel: Statistical errors; colors coincide with the top panel. The inset shows the statistical error relative to the error using nearest-integer resampling. Away from criticality, the curves differ significantly whereas around $\beta_c\approx 0.44$ the choice of the resampling method appears to have little effect on the statistical error.}
    \label{fig:SW_specHeat}
\end{figure}

\section{Numerical observations}
\label{sec:firstNumericalObservations}
In the following, we present numerical results for the benchmarking quantities introduced above. All data in this section were obtained through PA simulations using a target population size $R = 20\,000$, the annealing schedule $\beta_i = i / 300$ with $i\in\{0,\dots,300\}$, $\theta=5$ MCMC steps at each (inverse) temperature (except at $\beta_0 = 0$) and a linear system size $L=64$. For each resampling method independent simulations were run and repeated for $5\,000$ different random number seeds. Throughout this section, we visualize data sets by lines connecting \emph{all} points in the set. Additionally, a small subset of the data points is shown by points (with error bars when available).  Some preliminary results from these runs were presented in Ref.~\cite{Gessert2021}.

\begin{figure}
    \includegraphics[width=0.9\columnwidth]{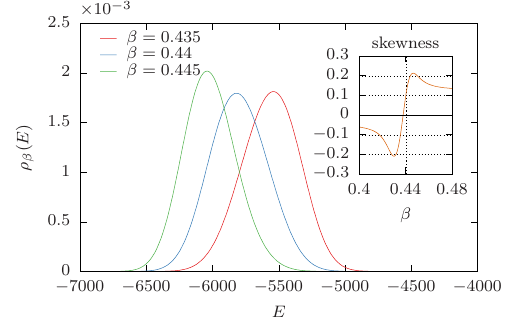}
    \caption{Exact energy distribution close to the inverse critical temperature $\beta_c \approx 0.44$ for the $64^2$ Ising model. Above (below) $\beta_c$ the histogram is skewed toward higher (lower) energies whereas very close to $\beta_c$ the distribution is symmetric. The skewness of the distribution as a function of $\beta$ is depicted in the inset which clearly shows a change in sign around $\beta_c$.}
    \label{fig:2DIM_skew}
\end{figure}

Figure~\ref{fig:SW_specHeat} shows systematic and statistical errors of the specific heat. Contrary to common practice, we measure \emph{before} the equilibration routine (i.e., immediately after resampling) as this is where the resampling-method dependent signal should be strongest. The use of this protocol is motivated by the fact that in target problems of PA such as models with complex free-energy landscapes MCMC equilibration routines are not as efficient, such that the effects of the choice of resampling scheme will be more prominent there. If we were to measure after the equilibration routine (as is usually done), then all errors would be indistinguishable except at the critical temperature due to the short autocorrelation time of the off-critical Ising model (not shown). We use the available exact solution~\cite{Kaufmann1949} to estimate systematic errors, and the standard deviation of independent runs for the statistical error. The most prominent feature in both types of error are two spikes around criticality, cf.\ Fig.~\ref{fig:SW_specHeat}. It is seen that all methods produce roughly comparable systematic and statistical errors, although multinomial and Poisson resampling have slightly stronger bias in the vicinity of the critical temperature and a substantially higher statistical error throughout. The inset in the lower panel shows all (statistical) errors relative to those for the nearest-integer method. It can be clearly seen that apart from systematic resampling all methods produce consistently larger errors than nearest-integer resampling.

The double peak of the statistical error and particularly the S-shape of 
the systematic error (which was previously reported by two of us, see Fig.~3 in Ref.~\cite{Ebert2022}) is readily understood through a change in sign of the skewness of the (exact) energy distribution (see Fig.~\ref{fig:2DIM_skew}): Below (above) the critical temperature the energy histogram is skewed toward higher (lower) energies. Consequentially, as low energies tend to be undersampled and high energies oversampled, the actual spread of the energy histogram and thus the specific heat is overestimated (underestimated). 



\begin{figure}
        \centering
        \includegraphics[width=0.9\columnwidth]{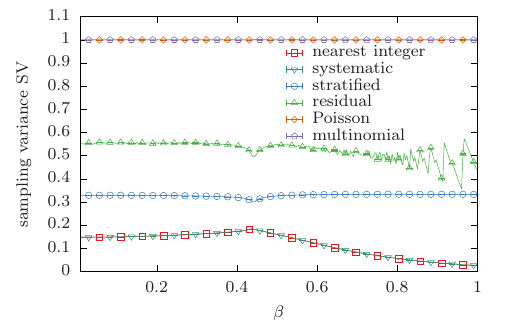}
        \caption{Sampling variance as a function of temperature for different resampling methods. Poisson and multinomial (respectively, nearest-integer and systematic) resampling have identical sampling variance. The parameters are as stated in the caption of Fig.~\ref{fig:SW_specHeat}. 
        }
        \label{fig:SW_Nivar}
    \end{figure}

As a measure of how much noise enters the setup through resampling, we consider the sampling variance~$\SV$ (see Fig.~\ref{fig:SW_Nivar}) which depends directly on the resampling method of choice and thus shows the strongest difference for the methods. Within the measurement accuracy multinomial and Poisson resampling have equal sampling variance. In fact, in both cases we measure a value of $1.0$ irrespective of temperature (for reasons that will become apparent in Sec.~\ref{sec:AsymptoticEstimator}). By value they have the highest $\SV$ followed by residual, stratified, systematic and nearest-integer resampling in that order. Systematic and nearest-integer resampling also coincide within the given accuracy. Note that this order of methods is the same as that observed in the inset in the lower panel of Fig.~\ref{fig:SW_specHeat}. The jumps in the $\SV$ of residual resampling can be explained by a discontinuity of the second moment of the probability distribution of $r_k$ for that particular method at $\tau_k = 1$ (for further detail see Appendix~\ref{app:residualResampling}).

\begin{figure}
    \centering
    \includegraphics[width=0.9\columnwidth]{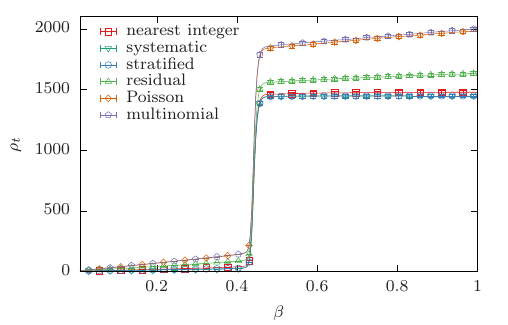}
    \includegraphics[width=0.9\columnwidth]{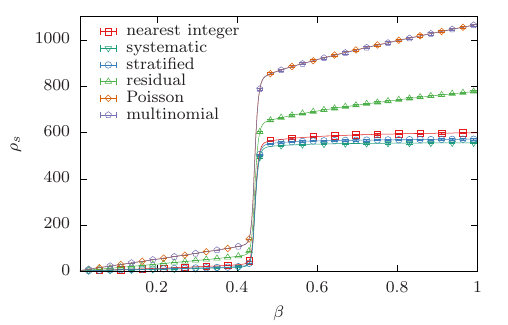}
    \includegraphics[width=0.9\columnwidth]{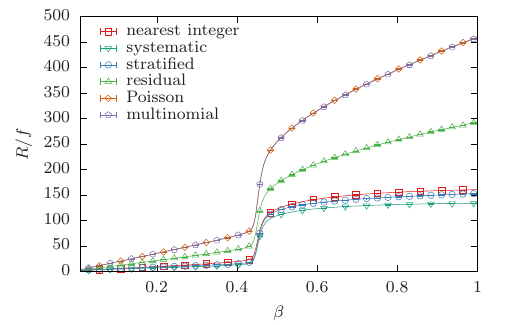}
    \caption{The family quantities $\rho_t$, $\rho_s$, and $f$ as a function of inverse temperature $\beta$ for the resampling methods studied ($R=20000, \theta=5, \Delta\beta=1/300$). Top panel: the replica-averaged family size $\rho_t$. Middle panel: the entropic family size $\rho_s$. Bottom panel: the plain average family size $R/f$. 
    }
    \label{fig:SW_familyQu}
\end{figure}

Our measurements for the family quantities introduced by Wang \emph{et al.}~\cite{Wang2015}, $\rho_t$, $\rho_s$ and $R/f$, are shown in Fig.~\ref{fig:SW_familyQu}. All quantities are monotonously increasing, with a rapid increase in the vicinity of the critical temperature. Larger families typically lead to a higher degree of correlation within the population and thus indicate worse statistics. Again multinomial and Poisson resampling perform worst, followed by residual resampling and then the remaining methods which produce similar results, particularly regarding the replica-averaged family size $\rho_t$.


\begin{figure}
    \centering
    \includegraphics[width=0.9\columnwidth]{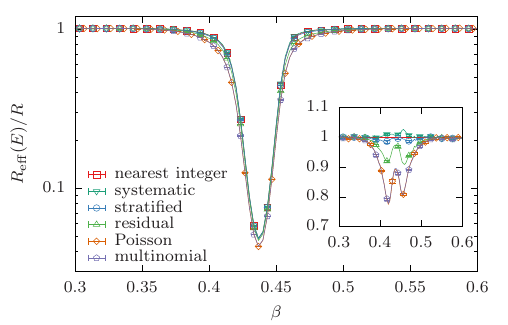}
    \includegraphics[width=0.9\columnwidth]{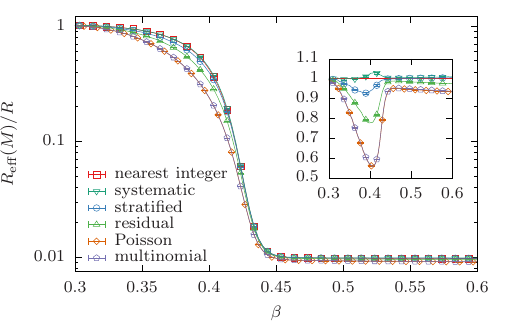}
    \caption{Effective population sizes $R_\text{eff}(E)$ (upper panel) and $R_\text{eff}(M)$ (lower panel) in units of $R$. The insets show the same quantities relative to the value observed with nearest-integer resampling ($R=20000, \theta=5, \Delta\beta=1/300$).}
    \label{fig:corrQu}
\end{figure}

Figure~\ref{fig:corrQu} shows the effective population sizes $R_\text{eff}(E)$ and $R_\text{eff}(M)$, where $E$ is the (extensive) internal energy of each configuration $\{\sigma_j\}$ [as given by Eq.~(\ref{eq:Hamiltonian})] and $M$ is the magnetization $M = \sum_j \sigma_j$. Both quantities are equal to the population size~$R$ at $\beta=0$ and show a dip at the inverse critical temperature. Thanks to the decorrelating nature of the MCMC updates $R_\text{eff}(E)$ recovers to the size of the population within the ordered phase, whereas $R_\text{eff}(M)$ remains low due to dynamic ergodicity breaking~\cite{Weigel2021}. In each panel the inset shows the effective population sizes compared to nearest-integer resampling. This again demonstrates that multinomial and Poisson resampling perform equally within the accuracy of our data and poorly as compared to the other methods. Nearest-integer and systematic resampling again show similarly good performance followed by stratified and residual resampling in that order. Also note that the methods which coincide in sampling variance, i.e., multinomial and Poisson resampling as well as systematic and nearest-integer sampling, respectively, also perform very similarly in all other benchmarking quantities. This indicates that the sampling variance captures the quality of a resampling method very well.

Note that $R_\text{eff}(M)$ for multinomial and Poisson resampling do not reach the same value in the low-temperature regime as the other methods do, as can be seen in the inset of the lower panel. This shows that a poor resampling method can amplify the correlation due to ergodicity breaking. In fact, with increasing $\beta$ the ratio of $R_\text{eff}(M)$ for multinomial and Poisson resampling with $R_\text{eff}(M)$ for the nearest-integer method decreases which suggests that for deep anneals well below a transition temperature these two methods are unfavorable. Similarly, the $R_\text{eff}(M)$ ratio for the residual resampling method also shows a decline in the ordered phase. We expect this observation to be of significance as ergodicity breaking is a common phenomenon in glassy systems. Hence, when studying such problems the multinomial, Poisson, and residual resampling methods should best be avoided. 

\section{Asymptotic analysis}

We now turn to an analysis of the asymptotic behavior of resampling methods for PA simulations of the Ising model, in particular considering the limits of $\theta\to\infty$ and $R\to\infty$.

\label{sec:thetaInfLimit}
\subsection{Exact sampling simulations}




From this point onward, we replace the Metropolis sampling of state space by an exact sampling of \emph{energies}. This is possible as the energetic density of states, $g(E)$, of the square-lattice Ising model can be calculated exactly for small to moderate system sizes~\cite{Beale1996}. Note that this is very different from the Propp-Wilson~\cite{ProppWilsonCFTP} method that achieves exact sampling of \emph{spin configurations}. We used the code provided in Ref.~\cite{Beale1996} to once calculate $g(E)$ for lattice sizes up to $L=128$, and stored $\ln g(E)$ in double precision for further use. From $\ln g(E)$, the probability of an energy $E$ at a given inverse temperature $\beta$ is easily obtained by computing

\begin{equation}
    P_\beta(E) = \frac{\exp\left(-\beta E + \ln g(E)\right)} {\sum_{E'} \exp\left(-\beta E' + \ln g(E')\right)}\,.
\end{equation}
In practice we draw from this distribution by using standard ``inverse transform sampling''.
\noindent The sum over all energies contains only $\textit{O}(L^2)$ terms and thus can be calculated very quickly. 
Clearly, combining exact sampling with population annealing is of no practical use as an actual simulation method. However, for us this artificial combination of PA and exact sampling is of theoretical interest as it isolates the effect of the choice of resampling method from the influence of imperfect MCMC equilibration. In this setting we can study the family quantities as well as the sampling variance whereas effective population sizes $R_\text{eff}(E)$ and $R_\text{eff}(M)$, as well as the statistical and systematic errors of observables become trivial.

\begin{figure}
    \centering
    \includegraphics[width=0.9\columnwidth]{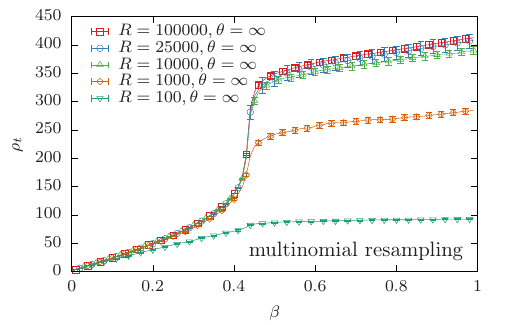}
    \includegraphics[width=0.9\columnwidth]{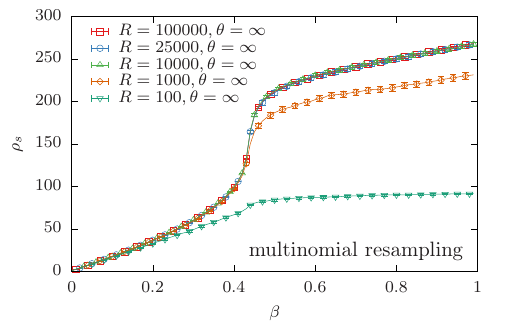}
    \includegraphics[width=0.9\columnwidth]{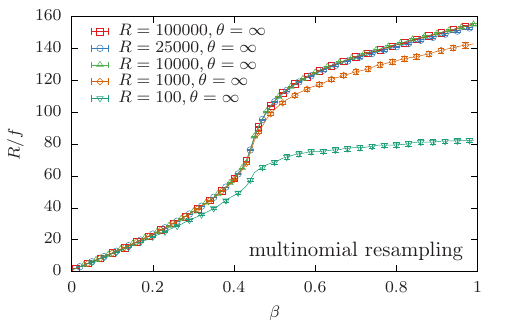}
    \caption{Family quantities $\rho_t$ (top), $\rho_s$ (middle), and $R/f$ (bottom) as a function of inverse temperature $\beta$ using inverse temperature step $\Delta \beta = 0.01$ and multinomial resampling. Lines connect all data points, but to improve readability only a small subset of the data points are shown explicitly.}
    \label{fig:familyQu_DS_popSizeEff}
\end{figure}

\begin{figure}
    \centering
    \includegraphics[width=0.9\columnwidth]{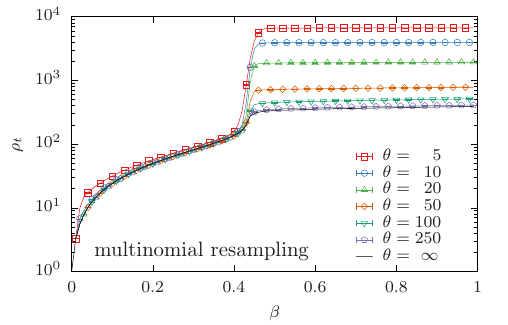}
    \caption{Family quantity $\rho_t$ for various choices of $\theta$. The population size was fixed to $10\,000$, $\Delta\beta$ was chosen as $0.01$ and multinomial resampling was used. Lines connect all data points, but only some are shown with explicit symbols.}
    \label{fig:familyQu_DS_thetaEff}
\end{figure}

Having eliminated the effect of imperfect equilibration, two key parameters besides the resampling method remain that also affect the PA simulation: the (target) population size $R$ and the chosen annealing schedule $\{\beta_i\}$. These are investigated in the present section.
Below in Sec.~\ref{sec:ResamplingCost} we introduce the notion of \emph{resampling cost} that allows us to compare the various resampling methods independent of the particular choice for $R$ and $\{\beta_i\}$.

The effect of the (target) population size on the family quantities $\rho_t$, $\rho_s$, and $R/f$ is illustrated in Fig.~\ref{fig:familyQu_DS_popSizeEff} for multinomial resampling. It can be seen that whenever $R$ is much larger than $\rho_t$, $\rho_s$, and $R/f$, respectively, the curves for different population sizes $R$ collapse. Hence, as long as the population size is large enough, it does not affect the family quantities; intuitively this is clear as then each family does not ``feel'' the finite (population) size and behaves as it would in the limit $R \rightarrow \infty$. Conversely, for small $R$ the asymptotic values of $\rho_t, \rho_s, $ and $R/f$ are underestimated.
For small $\beta$ the data for all population sizes studied agree and in order of ascending population size the individual data sets start deviating from the $R\rightarrow\infty$ case.

Note that compared to the results discussed in Sec.~\ref{sec:firstNumericalObservations} (see, for example, Fig.~\ref{fig:SW_familyQu}) values obtained for the family quantities here are significantly smaller. For such small~$\theta$, insufficient equilibration (which clearly is the case for $\theta=5$ at least close to the critical inverse temperature $\beta_c$) becomes the main driver of family growth resulting in much larger values for $\rho_t$, $\rho_s$ and $R / f$ in Fig.~\ref{fig:SW_familyQu}. The effect of $\theta$ is demonstrated in Fig.~\ref{fig:familyQu_DS_thetaEff} which shows measurements of $\rho_t$ for various choices of $\theta$. For $\beta < \beta_c$ almost all values for $\rho_t$ coincide and at $\beta_c$ all values for $\rho_t$ exhibit a strong increase: The smaller $\theta$ the stronger the increase in $\rho_t$ is.


\begin{figure*}
    \centering
    \includegraphics{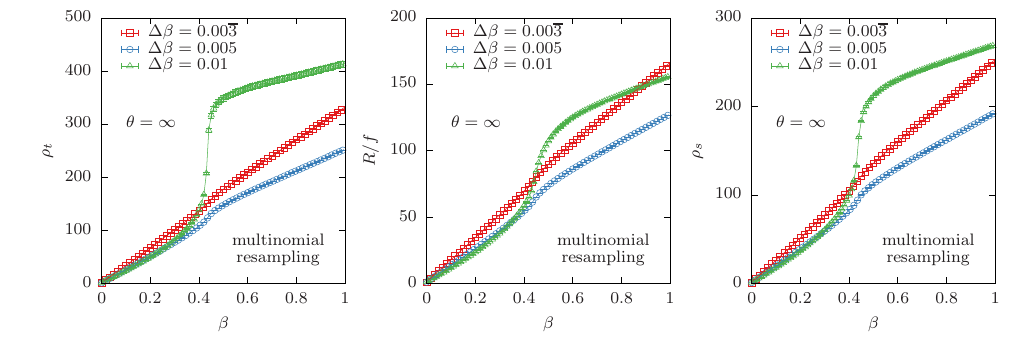}
    \caption{Family quantities $\rho_t$ (left), $\rho_s$ (middle), and $R/f$ (right) for population size $R=10^5$ at inverse temperature steps of $\Delta \beta = 0.00\overline{3}$, $0.005$, and $0.01$ and using multinomial resampling.}
    \label{fig:R100k_diffDbs}
\end{figure*}

Turning now to the annealing schedule, Fig.~\ref{fig:R100k_diffDbs} shows the family quantities for a fixed population size $R=10^5$ for various inverse temperature steps and again using multinomial resampling. The choice of $R=10^5$ is such that the effect of the population size being finite is negligible. At first glance, one might guess that smaller temperature steps would result in less family growth. However, Fig.~\ref{fig:R100k_diffDbs} shows a very different picture: While indeed (too) large temperature steps lead to a big increase in family size, also (too) small steps result in larger families than at intermediate temperature step sizes. This shows that too small temperature steps can in fact harm PA statistics instead of improving them. In Sec.~\ref{sec:ResamplingCost} this is made more explicit through the introduction of the notion of the resampling cost.

\subsection{Asymptotic estimator for the replica-averaged family size}
\label{sec:AsymptoticEstimator}
The standard method for calculating $\rho_t$ is
\begin{equation}
    \rho_t = R \sum_{k=1}^{R} \mathfrak{n}_k^2 = \frac 1 R \sum_{k=1}^{R} \mathfrak{N}_k^2\,,
    \label{equ:standard_rho_t}
\end{equation}
where $\mathfrak{n}_k$ (respectively, $\mathfrak{N}_k$) is the fraction (respectively, the number) of replicas descending from replica $k$ at $\beta_0=0$. As we show in Appendix~\ref{app:ProofImprEstRhoT}, when all $r_k$'s are i.i.d., then
\begin{equation}
    {\rho}_t^{(i)} \approx \rho_t^{(i-1)} + \sigma^2\left(r_k^{(i)}\right)\,. \label{equ:imprEstRhoT}
\end{equation}
This relation is noteworthy as it means that $\rho_t$ is expected to increase by $\Delta\rho_t \equiv \sigma^2(r_k^{(i)})$ irrespective of the previous value of $\rho_t$. Note that the estimator resulting from Eq.~(\ref{equ:imprEstRhoT}) is only correct if all the $r_k$'s are i.i.d. Generally, this is not the case as replicas may be strongly correlated within their families. However, in the limit $\theta\rightarrow\infty$ this assumption becomes true for most of the methods discussed here and approximately correct for all of them. 
See Appendix~\ref{app:ProofImprEstRhoT} for more detail.

Using the law of total variance, $\sigma^2\left(r_k^{(i)}\right)$ can be expressed as
\begin{equation}
    \Delta\rho_t = \sigma^2\left(r_k^{(i)}\right) = \sigma^2\left(\tau_k^{(i)}\right) + \SV(\beta,\Delta \beta)\,,
    \label{equ:riVariance}
\end{equation}
with $\tau_k^{(i)}$ being the expected number of copies of a replica and $\SV(\beta,\Delta \beta)$ the expected sampling variance when resampling from inverse temperature $\beta = \beta_{i-1}$ to ${\beta + \Delta \beta} = \beta_i$. Again, see Appendix~\ref{app:ProofImprEstRhoT} for a derivation of this relation. 

Note that this is not the first time $\rho_t$ has been studied in the MCMC-equilibrated regime. Reference~\cite{Amey2018} found that the increase in $\rho_t$ is approximately equal to $2\epsilon$ (cf. Eq.~(42) in Ref.~\cite{Amey2018}), where $\epsilon$ is the proportion of the population deleted during resampling, the so called culling fraction. This agrees with numerical results at high temperatures. Their calculation uses nearest-integer resampling and approximates the number of resampling steps $k$ to be a continuous ``time'' variable, thus assuming $\sigma^2(\tau_k^{(i)})=0$. In the limit of zero weight variance, and using nearest-integer resampling, it is easy to show that our derived formula agrees with Ref.~\cite{Amey2018}. However, Eq.~(\ref{equ:riVariance}) here generalizes to other resampling methods and nonzero $\sigma^2(\tau_k^{(i)})$.

The first term on the right-hand side of Eq.~(\ref{equ:riVariance}), the variance of the Boltzmann weights (which we will refer to as the \emph{weight variance}), is independent of the resampling method whereas the second term, the \emph{sampling variance}, is strongly resampling-method specific.
Roughly, these two terms can be understood as the two driving forces in the evolution of population quantities in the well-equilibrated regime. In the limit of very small $\Delta \beta$ the population is noise-driven (sampling variance dominates weight variance) and in the limit of large $\Delta \beta$ it is weight-driven. This explains three observations in Fig.~\ref{fig:R100k_diffDbs}: (i) In the case of large $\Delta\beta$, $\rho_t$ shows a strong increase near criticality due to a large weight variance. (ii) At small $\Delta\beta$, $\rho_t$ follows nearly a straight line as the weight variance is negligible compared to the constant sampling variance (dependent on the resampling method, see below)~\footnote{Note that this limit is different from the one in which Eq. (42) of Ref.~\cite{Amey2018} was derived.}. (iii) Last, this is the reason why $\rho_t$ is minimal for an intermediate step size. The inverse temperature step that achieves minimal average family size will in general depend both on the sampling variance and on the weight distribution. 

\begin{figure}[h]
    \centering
    \includegraphics[width=0.9\columnwidth]{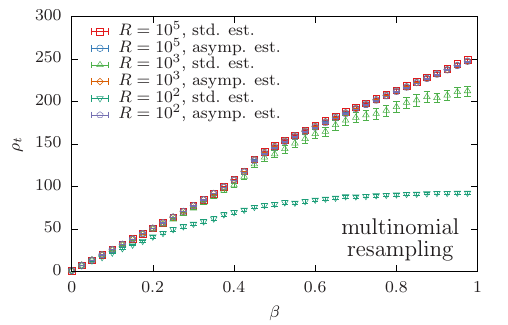}
    \includegraphics[width=0.9\columnwidth]{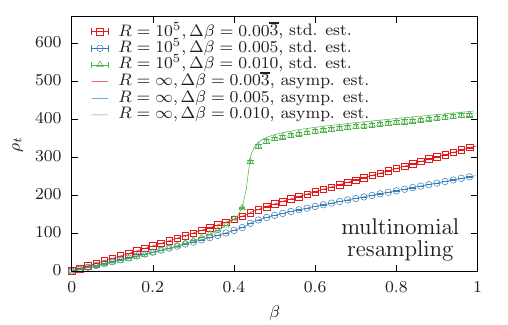}
    \caption{Comparison of $\rho_t$ measured with the standard estimator \eqref{equ:standard_rho_t} and with the asymptotic one \eqref{equ:imprEstRhoT}. Top: Comparison of different population sizes. Bottom: Comparison of different annealing schedules.}
    \label{fig:GE_imprEstRhoT}
\end{figure}

The variance of $r_k^{(i)}$ is straightforward to measure and thus $\rho_t$ can be estimated. The upper panel of Fig.~\ref{fig:GE_imprEstRhoT} shows the estimates for $\rho_t$ for various population sizes using the standard estimator \eqref{equ:standard_rho_t} as well as the asymptotic one \eqref{equ:imprEstRhoT}. It can be seen that both are in good agreement as long as the population size is much larger than $\rho_t$. The lower panel shows $\rho_t$ for various choices of $\Delta\beta$ (see also the left panel in Fig.~\ref{fig:R100k_diffDbs}), once using the standard estimator and $R=10^5$ and once using the asymptotic estimator and $R=\infty$ (details see below). As the population size is chosen large enough here, the agreement of the standard and the asymptotic estimator is very good. 

The reason we use the asymptotic estimator, is that $\lim_{R\rightarrow\infty}\sigma^2(r_k^{(i)})$ can be calculated [using Eq.~(\ref{equ:riVariance})] and thus $\rho_t$ can be evaluated [using Eq.~(\ref{equ:imprEstRhoT})] in the limit $R\rightarrow\infty$ which otherwise is not possible.
%
When resampling a population from inverse temperature $\beta$ to $\beta+\Delta\beta$, $\tau$ for a replica with energy $E$ is given by
\begin{equation}
    \tau(\beta,\Delta\beta,E) = e^{-\Delta\beta E} \frac{Z(\beta)}{Z(\beta + \Delta \beta)}\,. \label{equ:tauFromEnergy}
\end{equation}
In Eq.~(\ref{equ:tauFromEnergy}) we deliberately omitted a factor of $R/R_i$, which for population-size preserving methods equals one and otherwise approaches one as $R\rightarrow\infty$.
As above equation is a one-to-one mapping from $E$ to $\tau$ for fixed $\beta$ and $\Delta\beta$ and since the energy distribution in the $d=2$ Ising model is exactly known, the distribution of the $\tau$'s can easily be obtained. This allows for straightforward computation of the weight variance, $\sigma^2(\tau)$.

As for the calculation of the second term in Eq.~(\ref{equ:riVariance}), the sampling variance $\SV(\beta,\Delta\beta)$, this will in principle depend on the full set of $\{\tau_k\}$ for the entire population of $R_i$ replicas. For example when using the systematic or stratified resampling method, the sampling variance of replica $m$ specifically depends on $\sum_{k=1}^{m-1}\tau_k$'s as well as $\tau_m$ itself. In the following we denote $\sv_m = \langle (\tau_m - r_m)^2 \rangle_{\{\tau_k\}}$ as the sampling variance of a specific replica $m$ which in principle also depends on the full set of $\{\tau_k\}$ and is related to the previously defined $\SV$ through Eq.~(\ref{equ:asymptotic-sampling-variance}) given below. We can approximate $\sv_m$ as a function of a single $\tau_m$ for the various methods as follows (and as in this approximation $\sv_m$ only depends on its own $\tau_m$ we omit the index $m$): 

For multinomial and Poisson one obtains
\begin{subequations}
\begin{equation}
    \sv^{\text{mult}} (\tau) =\sv^{\text{Poi}} (\tau) =  \tau\,, \label{equ:samplingVarianceMultPois}
\end{equation}
which in the case of multinomial resampling is only strictly true in the limit $R \rightarrow \infty$. Note that $\tau$ on average is always one and thus the averaged sampling variance for the two methods is temperature-independent (as could be seen in Fig.~\ref{fig:SW_Nivar} above). For residual resampling we find
\begin{equation}
\sv^\text{res}(\tau) = \tau - \lfloor \tau \rfloor \equiv \epsilon\,,
\end{equation}
where $\epsilon$ refers to the fractional part of $\tau$. The sampling variance for stratified resampling is given by 
\begin{equation}
\sv^\text{strat} (\tau) = \begin{cases}
        \frac 1 3\,, & \tau \geq 1 \\
        \left(\frac{\tau^2}{3} - \tau + 1\right) \tau\,,  & \tau < 1
    \end{cases}\,,
\end{equation}
and for systematic as well as nearest-integer resampling by
\begin{equation}
    \sv^\text{ni}(\tau) = \sv^\text{sys}(\tau) = \epsilon (1 - \epsilon)\,. \label{equ:samplingVarianceNISyst}
\end{equation}
\end{subequations}

\noindent For the derivations of these relations we refer to Appendix~\ref{app:samplingVarianceCalc}.

\begin{figure}
    \centering
    \includegraphics[width=0.9\columnwidth]{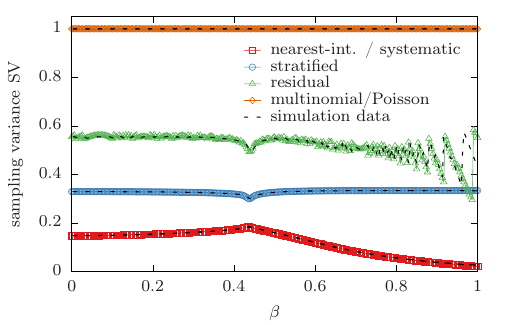}
    \caption{Sampling variance $\SV$ according to Eq.~\eqref{equ:asymptotic-sampling-variance} as function of the inverse temperature for nearest-integer/systematic, stratified and multinomial/Poisson resampling. $\Delta\beta = 1/300$.
    Dashed lines show the simulation data from Fig.~\ref{fig:SW_Nivar}.}
    \label{fig:GE_samplingVariance}
\end{figure}

From these, the averaged sampling variance for a given $\beta$ and $\Delta\beta$  can be calculated by summation over all possible energies, i.e.,
\begin{equation}
    \SV (\beta,\Delta\beta) = \sum_E p_{\beta_i}(E) \sv^m\left(\tau\left(\beta,\Delta \beta,E\right)\right)\,,
    \label{equ:asymptotic-sampling-variance}
\end{equation}
where the superscript $m\in \{\text{mult,Poi,res,strat,ni,sys}\}$ refers to the chosen resampling method.
As can be seen in Fig.~\ref{fig:GE_samplingVariance}, the $\SV$ calculated in this way is in very good agreement with values previously obtained through MCMC simulations. Here, the inverse temperature step was chosen as $\Delta\beta = 1/300$ to allow for comparison with the simulation data from Fig.~\ref{fig:SW_Nivar}.

Adding both terms in Eq.~\eqref{equ:riVariance} gives rise to an estimate for $\rho_t$ in the limit of $R\rightarrow\infty$ and for a given annealing schedule. The lower panel of Fig.~\ref{fig:GE_imprEstRhoT} shows quasiexact~\footnote{Quasi-exact refers to a large but finite sample obtained by exact sampling.} results from exact sampling PA compared to the calculation of $\rho_t$ in the limit $R\rightarrow\infty$ from the exact $\tau$-distribution which are very compatible.
Note that this calculation can be performed in time $\textit{O}(L^2)$ and in terms of wall-clock time the calculation of the quantities of interest in the double limit $\theta$ and $R \rightarrow \infty$ is near instantaneous.

\subsection{Notion of resampling cost}
\label{sec:ResamplingCost}
The goal of the considerations so far was to isolate the effect of the resampling method from that of other parameters, namely, the number of MCMC sweeps $\theta$, the population size $R$ and the annealing schedule $\{\beta_i\}$. The $\theta$-dependence was overcome by taking the limit $\theta\rightarrow\infty$ and similarly, the $R$-dependence was removed by taking the limit $R\rightarrow\infty$. The remaining dependence on the annealing schedule cannot be removed by a simple limit as $\rho_t$ at $\beta_i$ does not only depend on $\beta_i$ but on the entire schedule up to~$\beta_i$.


\begin{figure}
    \centering
    \includegraphics[width=0.9\columnwidth]{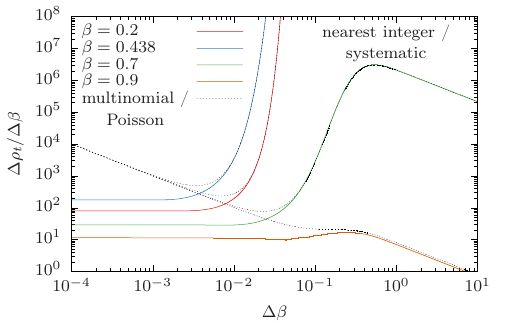}
    \includegraphics[width=0.9\columnwidth]{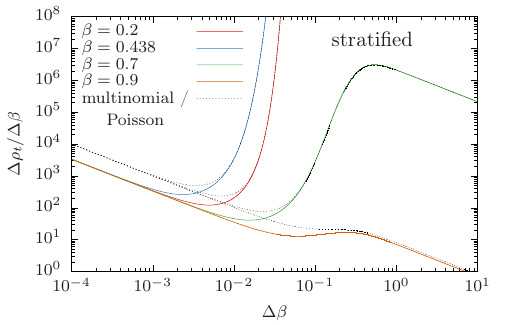}
    \includegraphics[width=0.9\columnwidth]{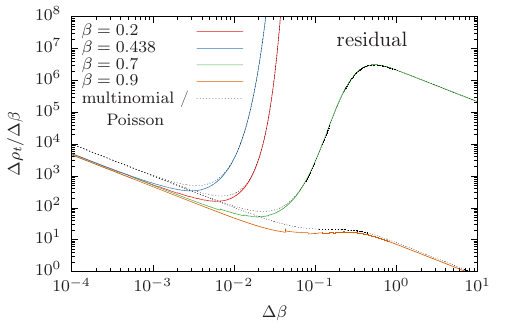}
    \caption{Increase of $\rho_t$ per step size $\Delta \beta$ (resampling cost) for various inverse temperatures obtained using different resampling methods. For large $\Delta \beta$ (limit of weight-driven) curves from different resampling methods coincide. For small $\Delta \beta$ (limit of noise-driven) curves differ significantly as noise-driven behavior depends predominately on the sampling variance $\SV$. Top panel: nearest-integer/systematic resampling. Middle panel: stratified resampling. Bottom panel: residual resampling.}
    \label{fig:resCostForDiffResamplingMethods}
\end{figure}

Instead, we note that in Eq.~(\ref{equ:imprEstRhoT}) the increase in $\rho_t$, namely $\Delta\rho_t = \rho_t(\beta_i) - \rho_t(\beta_{i-1})$ only depends on the two temperatures $\beta_{i-1}$ and $\beta_{i}$. This allows us to consider $\rho_t$ as a function of $\Delta\beta$ for various $\beta$. Clearly, $\Delta \rho_t$ (for any resampling method) will be the smallest when $\Delta\beta\rightarrow 0$. However, as we have pointed out above (see, e.g., Fig.~\ref{fig:GE_imprEstRhoT}) this will not result in the smallest possible \emph{final} $\rho_t$ at the inverse stopping temperature $\beta_s$. It is thus natural to define the resampling cost $\Delta\rho_t / \Delta \beta$ which is shown as a function of $\Delta\beta$ for various $\beta$ in Fig.~\ref{fig:resCostForDiffResamplingMethods}. The resampling cost can be understood as the cost per inverse-temperature step-width attributed to making a certain temperature step with a certain resampling method.

Studying this quantity reveals a very tight connection between resampling and the chosen temperature step. The two most interesting observations from the data shown in Fig.~\ref{fig:resCostForDiffResamplingMethods} can be made in the limit of very small and very large inverse temperature steps. In the limit of small (large) steps the second (first) term in Eq.~(\ref{equ:riVariance}) dominates and $\Delta \rho_t$ is noise-driven (weight-driven).
Hence, on the one hand, for large steps the resampling cost becomes independent of the chosen resampling method, which is illustrated through the collapse onto the dashed data set from multinomial resampling included in each panel. On the other hand, for very small temperature steps the resampling cost becomes independent of temperature and diverges as $1/\Delta\beta$ for all methods except nearest-integer and systematic resampling, where the resampling cost approaches a constant temperature-dependent value.


At very small $\Delta\beta$ the weight variance approaches zero and becomes negligible. When the weight variance is zero, all $\tau_k$'s are equal to one. In this case the sampling variance approaches a constant value which for
\begin{itemize}
    \item multinomial/Poisson resampling is one,
    \item for residual resampling is $\frac 1 2$,
    \item for stratified resampling is $\frac 1 3$, and
    \item for nearest-integer and systematic resampling equals zero.
\end{itemize}
These values can be found by taking the $\lim_{\tau\rightarrow 1} \sv(\tau)$ in Eqs.~(\ref{equ:samplingVarianceMultPois}) to (\ref{equ:samplingVarianceNISyst})~\footnote{When the left and the right limits differ (residual resampling), then the mean of the two values is taken, cf.\ also Appendix~\ref{app:residualResampling}.}. The nonzero limits for all but the nearest-integer and systematic resampling methods then lead to the divergent behavior observed in Fig.~\ref{fig:resCostForDiffResamplingMethods}.

\begin{figure}
    \centering
    \includegraphics[width=0.9\columnwidth]{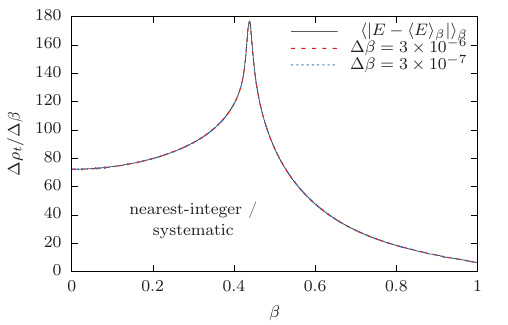}
    \caption{Resampling cost as a function of inverse temperature $\beta$ for the nearest-integer resampling method for very small temperature steps. In the limit of $\Delta\beta\rightarrow 0$, the resampling cost approaches the mean absolute deviation of the energy distribution, i.e., $\langle | E - \langle E \rangle_\beta | \rangle_\beta$.}
    \label{fig:GE_Tau2vsBetaNearestInt}
\end{figure}

The constant resampling cost for nearest-integer and systematic resampling can be understood by Taylor expanding $\tau_\beta(\Delta\beta,E') = e^{-\Delta\beta E'} Z(\beta) / Z(\beta + \Delta \beta)$ for small $\Delta\beta$ which up to first order in $\Delta\beta$ yields
\begin{equation}
    \tau(\beta,\Delta\beta,E') = 1 + \left(\langle E \rangle_\beta - E'\right) \Delta \beta + \textit{O}(\Delta \beta^2)\,,
\end{equation}
where $\langle E \rangle_\beta$ is the canonical average of the energy at inverse temperature $\beta$. Observing that for nearest-integer and systematic resampling the sampling variance [see Eq.~(\ref{equ:samplingVarianceNISyst})] in the limit $\tau \rightarrow 1$ approaches $|1 - \tau|$, one obtains
\begin{equation}
\SV(\beta,\Delta\beta) = \langle | E - \langle E \rangle_\beta | \rangle_\beta \Delta\beta + \textit{O}(\Delta \beta^2)\,.
\end{equation}
Here, $\langle | E - \langle E \rangle_\beta | \rangle_\beta$ is the mean absolute deviation (MAD) of the energy distribution, which is the temperature-dependent constant observed in the top panel of Fig.~\ref{fig:resCostForDiffResamplingMethods}. Figure~\ref{fig:GE_Tau2vsBetaNearestInt} shows that for very small $\Delta\beta$ indeed the values for $\Delta\rho_t/\Delta\beta$ and the MAD fall on top of each other. Note that this calculation was carried out in the limit of $R\rightarrow\infty$. When using the nearest-integer resampling method we expect a nonzero sampling variance of the order of $1/R$ as $\Delta \beta \rightarrow 0$. While theoretically this leads to an increase in $\SV$ as $\Delta \beta \rightarrow 0$ for finite population sizes, in any practical setting all other contributions to $\SV$ will be much larger than $1/R$ such that this effect is negligible.

Furthermore, the crossover point between the noise-driven and weight-driven regimes is strongly temperature-dependent as can be best seen by looking at the position of the minimum of the resampling cost for the remaining methods. Close to the critical temperature the weight-driven regime begins at much smaller $\Delta\beta$ whereas away from criticality even considerably large $\Delta\beta$ are within the noise-driven regime. Thus, if we fix $\Delta\beta$, then we expect a strong dependence on the resampling method away from criticality and almost no dependence around $\beta_c$ which is in very good agreement with our experimental observations in Sec.~\ref{sec:firstNumericalObservations} (see, for example, Fig.~\ref{fig:SW_specHeat}, in particular the inset in the lower panel).

\subsection{Optimal inverse temperature steps}
For all resampling methods the resampling cost becomes rather large when the chosen temperature step is (too) large. Further, for all methods but the nearest-integer and systematic ones the resampling cost also is quite large for (too) small inverse temperature steps. This naturally raises the question which temperature step to choose. In the following, we will demonstrate for the example of multinomial resampling how $\rho_t$ can be reduced by using an adaptive annealing schedule~\footnote{Note that the computational cost attributed to smaller or larger temperature steps is \emph{not} taken into account in the notion of resampling cost. This is particularly true as $\Delta\beta\rightarrow 0$.}.

\begin{figure}
    \centering
    \includegraphics[width=0.9\columnwidth]{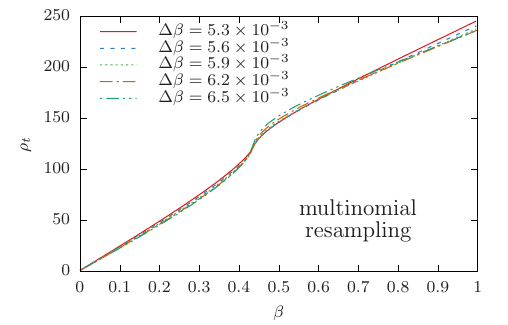}
    \includegraphics[width=0.9\columnwidth]{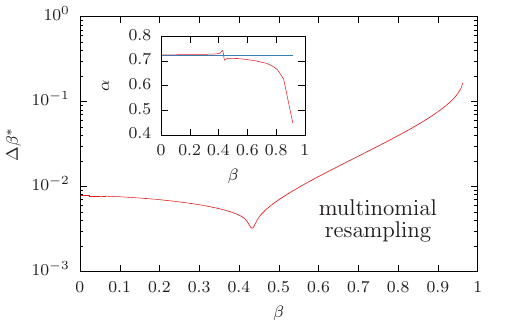}
    \includegraphics[width=0.9\columnwidth]{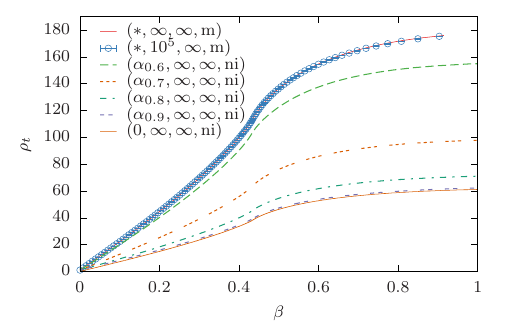}
    \caption{Optimization of the annealing schedule to minimize $\rho_t$ for the case of multinomial resampling. Top panel: $\rho_t$ for various linear schedules. Middle panel: Inverse temperature step $\Delta\beta^{*}$ at which the resampling cost is minimal. The inset shows the histogram overlap $\alpha(\beta,\beta+\Delta\beta)$~\cite{Barash2017} for the respective annealing schedule as well as $\alpha^*=1-\operatorname{erf}(1/4)$. Bottom panel: $\rho_t$ measured using different annealing schedules and resampling methods. The key encodes annealing schedule, population size, number of sweeps at each temperature and the resampling method as a four-tuple.}
    \label{fig:optimalAnnealingScheduleMultinomial}
\end{figure}

First, we attempt to find an optimal annealing schedule with a constant $\Delta\beta$. As is seen in the top panel of Fig.~\ref{fig:optimalAnnealingScheduleMultinomial}, the order of the lines changes with temperature, which suggests that the step should be chosen adaptively to minimize $\rho_t$. For example, $\Delta\beta = 6.5 \times 10^{-3}$ leads to the highest $\rho_t$ around $\beta=0.5$ and has one of the lowest $\rho_t$'s for $\beta \gtrsim 0.9$.

To minimize $\rho_t$ we can find the temperature step that results in a minimum increase in $\rho_t$ as a function of inverse temperature (see middle panel in Fig.~\ref{fig:optimalAnnealingScheduleMultinomial}). This is done by using the golden-section search method to find the $\Delta \beta$ at which the resampling cost is minimal. The optimal inverse temperature step $\Delta\beta^{*}$ obtained in this way varies over almost two orders of magnitude which shows that any linear annealing schedule gives sub-optimal results. $\Delta\beta^{*}$ is small (respectively, large) at (respectively, away from) the critical temperature suggesting temperature steps should be chosen smaller around the critical temperature than away from it. 

Starting at $\beta_0=0$, we choose the annealing schedule through a simple greedy strategy, i.e., $\beta_{i} = \beta_{i-1} + \Delta\beta^{*}(\beta_{i-1})$. The resulting measurement for $\rho_t$ corresponds to the datasets $(*,\infty,\infty,\mathrm{m})$ and $(*,10^5,\infty,\mathrm{m})$ in the bottom panel of Fig.~\ref{fig:optimalAnnealingScheduleMultinomial}. These are significantly lower in value than any $\rho_t$ obtained through schedules with constant $\Delta\beta$ (see top panel). The four entries of the tuples $(\mathcal{B},R,\theta,\mathcal{R})$ describe the simulation protocol used, where $\mathcal{B}$ stands for the annealing schedule and $\mathcal{R}$ represents the resampling method.

Choosing temperature steps adaptively has been previously suggested in Refs.~\cite{Amey2018,Barash2017}. In Ref.~\cite{Barash2017} the next temperature is chosen such that the estimated overlap of the Boltzmann distribution of the old and the new temperature is within a target interval. We have measured the histogram overlap with our annealing schedule and found (for the two-dimensional Ising model and using multinomial resampling) that it is close to $0.7$ (except at very low temperatures) but clearly not constant, see the inset in the middle panel of Fig.~\ref{fig:optimalAnnealingScheduleMultinomial}. Note that this result strongly depends on the chosen resampling method. For a method with lower sampling variance such as stratified resampling we expect this value to be higher. One can show that when all target distributions are Gaussian with the same variance, then the $\Delta\beta^{*}$ that minimizes $\Delta\rho_t$ will give rise to the overlap
\begin{equation}
\alpha^{*} = 1 - \textrm{erf}(1/4) \approx 0.72367\dots\,,
\end{equation}
where $\textrm{erf}$ is the error function. As is shown in the inset, for small $\beta$ the measured overlap coincides with $\alpha^{*}$.

Further, this consideration does not apply to nearest-integer and systematic resampling as the resampling cost in these cases approaches a constant value as $\Delta\beta\rightarrow 0$. For these methods, a linear schedule with small enough $\Delta\beta$ will give a final $\rho_t$ close to $\int_{\beta_0}^{\beta_s} \langle |E -\langle E \rangle_\beta |\rangle_\beta \,\mathrm{d} \beta$, which corresponds to the $(0,\infty,\infty,\mathrm{ni})$ dataset in the bottom panel of Fig.~\ref{fig:optimalAnnealingScheduleMultinomial} and which is well below $\rho_t$ obtained with multinomial resampling. As too small temperature steps might still result in higher computational cost~\cite{Barash2017}, it should nonetheless be beneficial to use adaptive temperature steps when using nearest-integer or systematic resampling. The datasets $(\alpha_x,\infty,\infty,\mathrm{ni})$ show the resulting $\rho_t(\beta)$ when using nearest integer or systematic resampling with a target overlap-parameter equal to $x$. As can be seen, when using the nearest-integer resampling method an overlap as low as $0.6$ still outperforms multinomial resampling. Increasing $x$ further reduces $\rho_t$. Note that in case one chooses an overlap of $0.9$ for this setup one achieves a value of $\rho_t$ that is only slightly above the minimal $\rho_t$ observed.


\subsection{System size dependence}

As outlined at the end of Sec.~\ref{sec:ResamplingCost}, we expect the resampling cost for nearest-integer resampling to be close to the MAD of the energy, independently of the model and system size considered. The MAD is bounded from above by the standard deviation, yielding the inequality
\begin{figure}
    \centering
    \includegraphics[width=0.9\columnwidth]{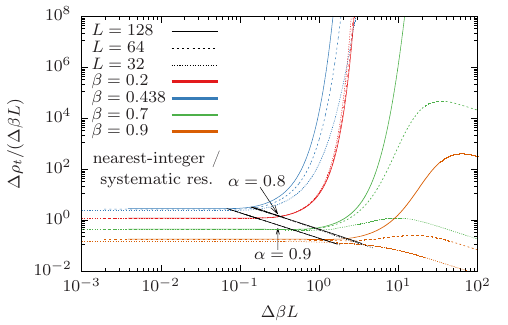}
    \caption{Resampling cost divided by linear system size $L$ as a function of inverse temperature step size scaled by the linear system size for nearest-integer and systematic resampling for different system sizes and temperatures. Line style (respectively, color) encodes system size (respectively, temperature). For small enough $\Delta \beta$ the resampling cost approaches the energy MAD which for the two-dimensional Ising model scales with $L$. The maximal temperature step at which the resampling cost is close to the MAD scales approximately with $1 / L$. See text for further detail.}
    \label{fig:Tau2vsDeltaBetaNI_diffSystemSizes}
\end{figure}
\begin{equation}
    \text{MAD} = \langle | E - \langle E \rangle_\beta | \rangle_\beta \leq \Big[\underbrace{\sigma^2(E)}_{C_V L^d / \beta^2}\Big]^{1/2} = C_V^{1/2} L^{d/2} / \beta. \label{equ:MAD_inequality}
\end{equation}
If we further assume the probability density $P_\beta(E)$ to be Gaussian, then the inequality can be replaced by an equality with $\sqrt{2/\pi}[\sigma^2(E)]^{1/2}$~\cite{Geary1935}.
%
It can be shown (see Appendix~\ref{app:ProofCvIntBound}) that the definite integral over $C_V^{1/2} / \beta$ has a system-size independent upper bound. Thus, we expect $\rho_t$ for nearest-integer resampling and with small enough temperature steps to behave as

\begin{equation}
\rho_t \propto L^{d / 2}\,.
\end{equation}
As we expect PA to perform poorly when $\rho_t$ is of the order of magnitude of the population size $R$, we require $\rho_t / R$ not to grow as the system size increases.
Thus, this relation is noteworthy as it motivates choosing the target population size proportional to $L^{d / 2}$ when using the PA algorithm for multiple system sizes (without any further assumptions on the underlying model). Note that this is not in contradiction with critical slowing down, as the number of sweeps $\theta$ near $T_c$ to achieve equilibration still has to scale as $O(L^z)$. In the case of the two-dimensional Ising model, $\rho_t$ hence scales proportional to $L$. In Fig.~\ref{fig:Tau2vsDeltaBetaNI_diffSystemSizes} we show the resampling cost divided by $L$. The collapse of the curves for small $\Delta\beta$ demonstrates that $\Delta\rho_t$ scales with $L$ for each temperature and thus agrees with $\rho_t \propto L^{d / 2}$.

To follow the scaling of the histogram overlap~\cite{Janke2007}, Ref.~\cite{Weigel2021} suggested to choose $\Delta\beta \propto 1 / L$. By choosing $\Delta\beta L $ as abscissa in Fig.~\ref{fig:Tau2vsDeltaBetaNI_diffSystemSizes} it can be seen that the inverse temperature at which the resampling cost starts deviating from the MAD also scales with $1 / L$, thus confirming $\Delta\beta \propto 1 / L$ to be a good choice.
Furthermore, in the figure it is highlighted where the histogram overlap $\alpha$ takes values $0.8$ and $0.9$ for 1\,000 inverse temperature points between $\beta=0$ and $1$ for the three system sizes.

\begin{figure}
    \centering
    \includegraphics[width=0.9\columnwidth]{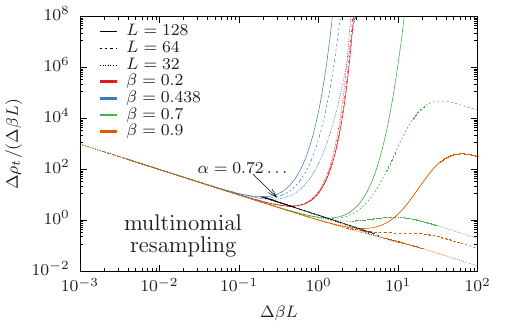}
    \includegraphics[width=0.9\columnwidth]{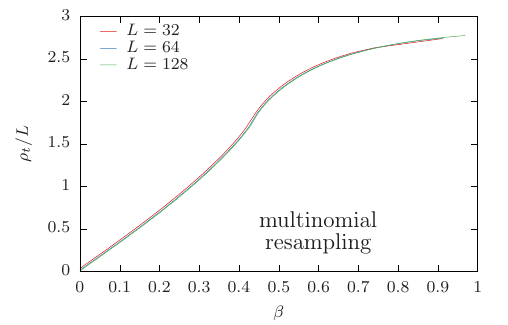}
    \caption{Adaptive temperature steps for different system sizes and multinomial resampling. Top panel: Resampling cost rescaled by the system size shows that the optimal temperature step $\Delta\beta^{*}$ behaves as $L^{-1}$. Bottom panel: $\rho_t$ as a function of $\beta$ using the optimized temperature steps scales proportional to $L$.}
    \label{fig:optimalAnnealingScheduleMultinomial_differentSystemSizes}
\end{figure}

Similar scaling is observed for other resampling methods as can be seen for the example of the multinomial resampling technique (see Fig.~\ref{fig:optimalAnnealingScheduleMultinomial_differentSystemSizes}). $\Delta\rho_t$ also scales with $L$ and the position of the optimum $\Delta\beta^{*}$ with $1/L$ (top panel). It is highlighted where $\alpha$ takes the value $\alpha^{*}\approx 0.72\dots$ which shows that indeed this value for $\alpha$ is close to the minimum of the resampling cost for the considered inverse temperatures and system sizes. In the bottom panel $\rho_t(\beta) / L$ for adaptive temperature steps is shown. From there one clearly sees that $\rho_t$ does scale as $L$. So, doubling the system size means that the number of temperature steps and the population size should be doubled as well (in $d=2$).

\subsection{Comparison to MCMC simulations}

\begin{figure}
    \centering
    \includegraphics[width=0.9\columnwidth]{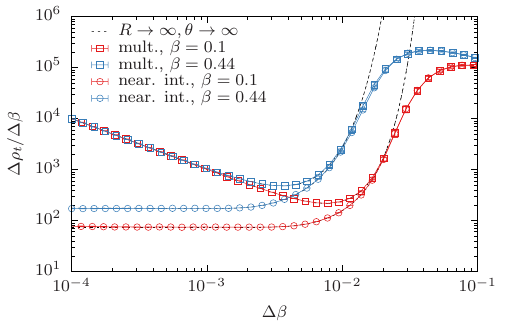}
    \includegraphics[width=0.9\columnwidth]{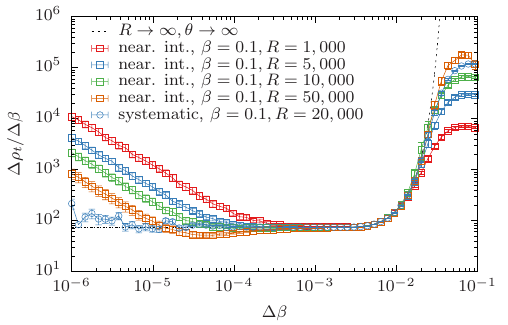}
    \caption{Resampling cost measurements on realistic standard PA simulations compared to the $R,\,\theta\rightarrow\infty$ limit. Upper panel: Multinomial resampling and nearest-integer resampling for two different temperatures. Lower panel: Nearest-integer and systematic resampling. At very small $\Delta\beta$ measurements deviate from the prediction, more strongly for nearest-integer resampling where population size fluctuations lead to a $\propto 1/\Delta\beta$ divergence (see main text). }
    \label{fig:resamplingCostInRealSimulations}
\end{figure}

In this last part, we once again return to more realistic population annealing simulations to probe how applicable our theoretical findings are to (close-to-equilibrium) MC simulations. Clearly, unlike the artificial simulations, they are not in perfect equilibrium and thus results will differ. Further, we expect this difference to be weak (respectively, strong) when close to (respectively, far from) equilibrium. As a consequence, the previously introduced resampling cost for small $\Delta\beta$ should not vary much in either setting whereas for large $\Delta \beta$ we anticipate strong differences.

Experimentally, it is not immediately clear how to measure $\Delta\rho_t / \Delta \beta$ at a given $\beta$. As the asymptotic estimator \eqref{equ:imprEstRhoT} for $\rho_t$ is only correct in the limit $\theta\rightarrow\infty$, we have to fall back to the standard estimator which explicitly depends on the population at $\beta$, raising the question of how to initialize the population at $\beta$ prior to the $\Delta\rho_t / \Delta \beta$ measurement. A number of options come to mind:

\begin{enumerate}
    \item For each $\Delta \beta$, run a PA simulation with this temperature step from $\beta_0 = 0$ to $\beta_i=\beta$.
    \item Create an uncorrelated population sample by quenching each replica (either from a hot or a cold start) for a very long time.
    \item Pick a temperature step $\widetilde{\Delta\beta}$ and initialize the population by running a simulation with $\widetilde{\Delta\beta}$ up until $\beta$.
\end{enumerate}

Best results are most likely obtained through option~1 but we have rejected this option as it is computationally prohibitively expensive, particularly for small $\Delta\beta$. Option~2 may work well but also produces uncorrelated replicas, which is rather unrealistic for PA simulations. Option~3 is straightforward to implement and produces realistic populations at inverse temperature $\beta$ although care must be taken of how to choose $\widetilde{\Delta\beta}$ which we set to $\widetilde{\Delta\beta} = 1/300$. This value we know from previous studies to work well on the considered model. An overview of the resampling cost for realistic simulations is presented in Fig.~\ref{fig:resamplingCostInRealSimulations}, where for each data set (each line) we ran 400 individual PA simulations. For the simulations initialized at $\beta=0.44$, we used $\theta=100$ sweeps at each temperature, and $\theta=10$ for the ones at $\beta=0.1$. As system size we used $L=64$.

The upper panel of Fig.~\ref{fig:resamplingCostInRealSimulations} shows MC results obtained with multinomial and nearest-integer resampling which are very compatible with the theoretical data up until $\Delta\beta \simeq 10^{-2}$ even when $\beta$ is close to the critical temperature. At very small $\Delta\beta$ we observe an unexpected divergence (see lower panel) in $\Delta\rho_t/\Delta\beta$ for nearest-integer resampling, which is strongly population-size dependent and which is absent when repeating the same measurement for systematic resampling. This is not in contradiction to our original expectation but rather an artifact of the way the population was initialized:
Following the procedure described above, the 400 initialized populations will vary in population size (dictated by the choice of $\widetilde{\Delta\beta}$).
The nonzero population-size fluctuation results in a nonzero increase of $\Delta\rho_t$ as $\Delta\beta\rightarrow 0$, which leads to the divergent behavior for $\Delta \rho_t / \Delta \beta$ observed in Fig.~\ref{fig:resamplingCostInRealSimulations}. The fluctuation of the population size is of the order of $\sqrt R$ which explains the strong dependence on $R$ in the plot. Without this fluctuation (albeit when using systematic resampling) this effect is absent.


As has become clear before, in a poorly equilibrated population $\rho_t$ is much larger than in the $\theta\rightarrow\infty$ limit, cf. Fig.~\ref{fig:familyQu_DS_thetaEff}. Thus, there ought to be a contribution not taken into account in the calculation of $\Delta\rho_t$ in the ${\theta\rightarrow\infty}$ limit, that leads to a sharp increase of $\rho_t$ at the inverse critical temperature $\beta_c \approx 0.44$. The crucial assumption that leads to the expressions in the $\theta\rightarrow\infty$ limit was that the weights of replicas are distributed independently, i.e., that the covariance of weights of replicas $i$ and $j$ (with $i \neq j$) is zero. This however, is not the case when $\theta$ is finite. In this case the covariance can roughly be estimated through the effective population size $R_\text{eff}(\tau)$ giving rise to an \emph{extra} contribution $\delta$ to $\Delta\rho_t$ that is 
\begin{equation}
\delta \sim \left(\frac {R}{R_\text{eff}(\tau)} -1\right)\,.\label{eq:extraRhoTcontrib}
\end{equation}
Note that when $R_\text{eff}(\tau) = R$ this contribution is zero (perfect equilibration), whereas it becomes large when the population is very correlated, i.e., when $R_\text{eff}(\tau) \ll R$. As $\Delta\beta$ and $\theta$ were chosen constant for each simulation in Fig.~\ref{fig:familyQu_DS_thetaEff}, the population is well-equilibrated even for small~$\theta$ except around criticality. Hence, Eq.~(\ref{eq:extraRhoTcontrib}) explains the jump at $\beta_c$ and away from criticality the lines are almost parallel.

\section{Conclusions}
\label{sec:conclusion}
Combining data from numerical simulations as well as theoretical considerations we provide strong evidence that the chosen resampling method in PA can have a significant effect on the quality of the obtained data. Out of all the methods we have considered, we find a range of results that suggest that using multinomial or Poisson resampling (which both have been used extensively in practice before) should be avoided. Both lead to higher systematic and statistical errors as well as worse values in all other considered benchmark quantities. Instead, nearest-integer and systematic resampling consistently outperform all other methods in almost all considered metrics and work equally well in the remaining comparisons.

Besides simply answering the question of which resampling method to choose, we further aimed to improve our understanding of the genuine effects of resampling on the simulation results. Replacing MCMC sweeps with exact sampling from the energy distribution (corresponding to $\theta\rightarrow\infty$) allowed us to systematically scan the parameter space of PA at  very little computational cost. What is more, however, it isolated the potential negative effects of resampling from the systematic error caused by imperfect equilibration due to finitely many MCMC sweeps.

In this setting we varied the population size $R$ while keeping all other parameters constant and found that when $R$ grows large, the family quantities smoothly converge in $R$, which is in agreement with previous reports~\cite{Wang2015,Weigel2021}. This further motivated us to consider PA in the double limit of perfect equilibration ($\theta\rightarrow\infty$) and infinite population sizes ($R\rightarrow\infty$), where we indeed see that the replica-averaged family size $\rho_t$ for large $R$ behaves in the same way as for $R\rightarrow\infty$.
%
We evaluated $\rho_t$ in this (double) limit by expressing it as the sum of the accumulated weight variances and sample variances [see Eq.~(\ref{equ:imprEstRhoT})]. Computationally this limit is even less demanding than the previous simple sampling approach as every setting can be evaluated by a single summation over $\textit{O}(L^2)$ terms which allowed us to thoroughly investigate the interplay between resampling and the chosen annealing schedule.

This interplay can best be observed by considering the increase in $\rho_t$ per inverse temperature step $\Delta\beta$ at a given temperature, which we dubbed the resampling cost. For all methods but nearest-integer and systematic resampling the resampling cost exhibits a minimum at an intermediate temperature step and diverges as $\Delta\beta\rightarrow 0$. This divergence is the reason why we observe worse statistics with very small temperature steps for most methods, which at first might have seemed counter-intuitive.

Furthermore, we show that by choosing temperature steps that minimize the resampling cost, even with multinomial and Poisson resampling acceptable results can be achieved, although nearest-integer and systematic resampling still yield smaller family growth. Comparing our adaptive schedule with the adaptive schedule resulting from fixing the histogram overlap~\cite{Weigel2021}, we observe that for multinomial resampling the target overlap should not be chosen much larger than $\approx 0.72$ as then $\rho_t$ will take sub-optimal values. As the resampling cost for systematic and nearest-integer resampling does not diverge in the small step limit, adaptive steps minimizing the resampling cost are not applicable to these two methods. Thus, a linear annealing schedule with small enough steps yields close-to-minimal family growth with these two methods. As for these two methods, we find that with a target overlap parameter of $\alpha \approx 0.8 \dots 0.9$ the replica-averaged family size $\rho_t$ takes close-to-minimal values, that is the value it would attain for $\alpha \rightarrow 1$.

As a last test in this artificial setting, we repeated the experiment for different linear system sizes $L$. In agreement with Ref.~\cite{Weigel2021} we find that for the two-dimensional Ising model the temperature step should be proportional to $1/L$, i.e., the position of the minimum (respectively, the length of the plateau for nearest-integer and systematic resampling) of the resampling cost scales as $1/L$. Most strikingly, we find that, assuming good equilibration and an appropriate annealing schedule, $\rho_t$ should scale as $L^{d/2}$ independent of the underlying model which gives rise to the simple rule of how large the population size $R$ should at least be when studying multiple system sizes of the same model.

Finally, we repeat some of these experiments with MCMC simulations to confirm that indeed close-to-equilibrium simulations behave similar to the idealized case. The agreement is surprisingly good although one should stress that the two-dimensional Ising model is easy to equilibrate as compared to spin glasses, which are notoriously hard to equilibrate and form one of the main applications of PA~\cite{Hukushima2003,Wang2015}. We find that in out-of-equilibrium PA simulations there is an extra contribution to the family growth that can be linked to the correlation within the population. Thus, we expect that by introducing a target effective population size $R_\text{eff}$ family growth may be well-controlled.

Although we set out to understand one aspect of PA, namely resampling, we can extract a few guiding principles from these studies regarding the question of how to choose $R$, $\theta$ and $\beta_i$ (for further rules of thumb of how to successfully implement a PA simulation we refer to Sec.~X in Ref.~\cite{Weigel2021}):

\begin{itemize}
    \item Use nearest-integer resampling. If a constant population size is required, then use systematic resampling. In particular, avoid the use of Poisson or multinomial resampling.
    \item Small enough temperature steps generally will give very good results when using one of the two preferred resampling methods. However, very small $\Delta\beta$ still come at a high computational cost and adaptive temperature steps will be more resource-efficient~\cite{Weigel2021}. Unless adaptive steps are used, $\Delta\beta$ should be scaled accordingly when studying multiple system sizes.
    \item When studying multiple linear system sizes $L$ of the same model, the population size should be scaled at least with $L^{d/2}\equiv \sqrt{N}$ as this is the expected behavior for $\rho_t$ in the well-equilibrated regime.
    \item Section~\ref{sec:thetaInfLimit} shows what to expect from well-equilibrated PA simulations. If you suspect poor equilibration, then increase $\theta$ if possible.

\end{itemize}

While we have discussed resampling in PA in rather general terms, not all possible variations are contained in this work. This includes resampling only at some of the temperature steps (as originally suggested by Hukushima and Iba~\cite{Hukushima2003}) or not resetting all weights after resampling~\cite{Doucet2001}, which could also be explored. Another interesting direction is to temporarily increase the population size before resampling by selecting multiple configurations per replica and then resample to the original population size as was recently proposed by Amey and Machta~\cite{Amey2021}. Last, when using PA on distributed parallel architectures resampling is directly linked to communication overhead. Thus, one potential area of focus is optimizing resampling for maximal parallel performance on these setups.

\begin{acknowledgments}
    The project was supported by the Deutsch-Französische Hochschule (DFH-UFA) through the Doctoral College ``$\mathbb{L}^4$'' under Grant No.\ CDFA-02-07. We further acknowledge support by the Leipzig Graduate School of Natural Sciences ``BuildMoNa''.
\end{acknowledgments}

\appendix
\section{Replica-averaged family size}
\label{app:RhoTName}
To elaborate on how the term ``mean-square family size'' for $\rho_t$ is misleading, consider a population of $R$ replicas and replica-averaged family size $\rho_t^{(R)}$. Now, if we were to copy every replica exactly twice, thus obtaining a new population of size $2R$ with $\rho_t^{(2R)}$, then we would expect the mean-square family size to quadruple. However, $\rho_t^{(2R)} = 2 \rho_t^{(R)}$.

We motivate the term ``replica-averaged family size'' as follows. Let $\mathfrak{o}_j$ be the index of the replica at $\beta_0$ from which replica $j$ originates. Then $j$ will be part of a family with $\mathfrak{n}_{\mathfrak{o}_j} R$ members. If we pick a replica uniformly at random (out of the entire population), then we expect its family size to be $\frac 1 R \sum_{j=1}^R \mathfrak{n}_{\mathfrak{o}_j} R$. Clearly, for each family of descendants from the original replica $k$ there are $R \mathfrak{n}_k$ contributions in this sum and thus we can rewrite it as

\begin{equation}
    \frac 1 R \sum_{j=1}^R \mathfrak{n}_{\mathfrak{o}_j} R = \frac 1 R \sum_{k=1}^R \mathfrak{n}_{k} R ~ \mathfrak{n}_{k} R =  R \sum_{k=1}^{R} \mathfrak{n}_k^2 \equiv \rho_t\,,
\end{equation}
    
\noindent
which coincides with the previous definition of $\rho_t$.
\section{Proof of Equation~(\ref{equ:imprEstRhoT})}
\label{app:ProofImprEstRhoT}
We set out to prove Eq.~(\ref{equ:imprEstRhoT}), i.e., $\rho_t^{(i)} \approx \rho_t^{(i-1)} + \sigma^2\left(r_k^{(i)}\right)$, under the assumption that all $r_k^{(i)}$ are i.i.d. We denote by $(*)$ when this assumption is used. In the limit of $\theta\rightarrow\infty$, $\tau_k^{(i)}$ are i.i.d. by design. When $r_k^{(i)}$ (for a single $k$) is a univariate random variable only dependent on $\tau_k^{(i)}$ (as is the case for all methods with variable population size considered here), the i.i.d. property of $r_k^{(i)}$ follows immediately. Special attention is required for the population-size preserving methods, see below.

The replica-averaged family size [see Eq.~(\ref{equ:rho_t})] is $\rho_t = R \sum_{k=1}^{R} \mathfrak{n}_k^2.$
When expressed in terms of the family size $\mathfrak{N}_k = R \mathfrak{n}_k$ it becomes
\begin{equation}
    \rho_t = \frac 1 R \sum_{k=1}^R \mathfrak{N}_k^2 = \overline{\mathfrak{N}_k^2} \sim \langle \mathfrak{N}_k^2 \rangle\,. \label{equ:rhoTitoFamilysize}
\end{equation}
Above equality shows that $\rho_t$ is the population average of $\mathfrak{N}_k^2$ and thus estimates $\langle \mathfrak{N}_k^2 \rangle$ which is closely related to the variance of the size of a family, $\mathfrak{N}_k$, i.e.,
\begin{equation}
    \texttt{Var}\left(\mathfrak{N}_k\right) = \langle \mathfrak{N}_k^2 \rangle - \langle \mathfrak{N}_k \rangle^2 = \langle \mathfrak{N}_k^2 \rangle - 1\,, \label{equ:varN}
\end{equation}
where $\langle \mathfrak{N}_k \rangle = 1$ because otherwise the (expected) population size would change throughout the anneal.

In the following a time superscript $(i)$ is added to all time-dependent quantities, where time corresponds to the number of resampling steps already carried out. Further, let $\mathcal{F}_k^{(i)}$ be the set of replica indices belonging to the family $k$ at time $i$. By design, this set contains $\mathfrak{N}_k^{(i)}$ elements, i.e., $\mathfrak{N}_k^{(i)} = |\mathcal{F}_k^{(i)}|$. Last, $r_k^{(i)}$ is the number of copies created of replica $k$ at the $i$th resampling step.
\begin{widetext}
From the setup of the algorithm (see Sec.~\ref{sec:algoPA}), the size of family $k$ at time $i+1$ is given by
\begin{equation}
\mathfrak{N}_k^{(i+1)} = \sum_{j \in \mathcal{F}_k^{(i)}} r_j^{(i+1)}\,.
\end{equation}
Carefully note that all terms in above equation are random variables. Fixing one initial replica $k$ and the corresponding set of family indices $\mathcal{F}_k^{(i)}$ gives rise to the conditional variance
\begin{align}
    \texttt{Var}\left(\mathfrak{N}_k^{(i+1)} | \mathcal{F}_k^{(i)}\right) &= \texttt{Var}\left(\sum_{j \in \mathcal{F}_k^{(i)}} r_j^{(i+1)}\bigg| \mathcal{F}_k^{(i)}\right) \stackrel{(*)}{=} \underbrace{|\mathcal{F}_k^{(i)}|}_{\mathfrak{N}_k^{(i)}} ~ \texttt{Var}\left(r_1^{(i+1)}\right)\,, \nonumber \\
    \Rightarrow \texttt{Var}\left(\mathfrak{N}_k^{(i+1)} | \mathfrak{N}_k^{(i)}\right) &= \mathfrak{N}_k^{(i)} ~ \texttt{Var}\left(r_1^{(i+1)}\right)\,, \label{eq:Nconditional}
\end{align}
where we have used the fact that $\texttt{Var}(\mathfrak{N}_k^{(i+1)}|\mathcal{F}_k^{(i)})$ only depends on $|\mathcal{F}_k^{(i)}|$ instead of $\mathcal{F}_k^{(i)}$.
\noindent
Using the law of total variance $\texttt{Var}\left(\mathfrak{N}_k^{(i+1)}\right)$ can be expressed as
\begin{equation}
\texttt{Var}\left(\mathfrak{N}_k^{(i+1)}\right) = \texttt{E}\left[\underbrace{\texttt{Var}\left(\mathfrak{N}_k^{(i+1)} | \mathfrak{N}_k^{(i)}\right)}_{\stackrel{(\ref{eq:Nconditional})}{=} \mathfrak{N}_k^{(i)} ~ \texttt{Var}\left(r_j^{(i+1)}\right)}\right] + \texttt{Var}\left[\underbrace{\texttt{E}\left(\mathfrak{N}_k^{(i+1)} | \mathfrak{N}_k^{(i)}\right)}_{= \mathfrak{N}_k^{(i)}}\right]\,.
\end{equation}
This further simplifies to 
\begin{equation}
\texttt{Var}\left(\mathfrak{N}_k^{(i+1)}\right) = \underbrace{\texttt{E}\left( \mathfrak{N}_k^{(i)}\right)}_{=1}  \texttt{Var}\left(r_j^{(i+1)}\right) + \texttt{Var}\left( \mathfrak{N}_k^{(i)}\right)\,.
\end{equation}
Plugging in Eq.~(\ref{equ:varN}) for $\texttt{Var}\left(\mathfrak{N}_k^{(i)}\right)$ and $\texttt{Var}\left(\mathfrak{N}_k^{(i+1)}\right)$ gives rise to 
\begin{equation}
\left\langle\left( \mathfrak{N}_k^{(i+1)}\right)^2\right\rangle = \texttt{Var}\left(r_j^{(i+1)}\right) + \left\langle\left( \mathfrak{N}_k^{(i)}\right)^2\right\rangle\,.
\end{equation}
Substituting $\rho_t^{(i)}$ and $\rho_t^{(i+1)}$ for $\langle (\mathfrak{N}_k^{(i)})^2 \rangle$  and $\langle (\mathfrak{N}_k^{(i+1)})^2 \rangle$ [using Eq.~(\ref{equ:rhoTitoFamilysize})] completes the proof and gives rise to Eq.~(\ref{equ:imprEstRhoT}).
Regarding population-size preserving methods, if all $r_j$ were i.i.d., then their sum would have to be a random variable, too. Thus, for these methods the $r_j$ cannot be i.i.d. In the case of multinomial, and thus residual resampling, in the limit $R \rightarrow \infty$ the correlation between any two $r_i$ and $r_j$ vanishes and therefore the $r_j$ can be assumed to be approximately i.i.d. Only for stratified and systematic resampling this is not the case, making Eq.~(\ref{equ:imprEstRhoT}) only approximately true due to nonvanishing spatial correlations (in replica space). The correlation in those two methods causes Eq.~(\ref{equ:imprEstRhoT}) to slightly overestimate $\rho_t$ as compared to the standard definition.

Last, Eq.~(\ref{equ:riVariance}) is found by using the law of total variance: $\texttt{Var}(r_j^{(i)})$ can be expressed as
\begin{equation}
\texttt{Var}\left(r_j^{(i)}\right) = \underbrace{\texttt{E}\left[\texttt{Var}\left(r_j^{(i)} | \tau_j^{(i)}\right) \right]}_{\SV(\beta_{i-1},\beta_{i-1}-\beta_{i})} + \texttt{Var}\left[\underbrace{\texttt{E}\left(r_j^{(i)} | \tau_j^{(i)}\right)}_{= \tau_j^{(i)}} \right]\,,
\end{equation}
which yields Eq.~(\ref{equ:riVariance}).

\end{widetext}

\section{Sampling variance calculations}
\label{app:samplingVarianceCalc}
For all following calculations $\tau$ denotes the expectation value of the number of descendants $r$ of an arbitrary population member at time of resampling. More precisely, $\tau$ and $r$ should carry a replica index $k$. Unless strictly necessary, however, $k$ will be omitted for better readability. Further, $\epsilon$ denotes the fractional part of $\tau$, i.e., $\epsilon = \tau - \lfloor \tau \rfloor$. 
\subsubsection*{Nearest-integer resampling}
$r$ is $\lfloor\tau\rfloor + 1$ with probability $\epsilon$ and $\lfloor\tau\rfloor$ with probability $1-\epsilon$. Thus, it is easy to see that $r-\tau$ is $1-\epsilon$ with probability $\epsilon$, and $-\epsilon$ with probability $1-\epsilon$. Hence the sampling variance is 
\begin{equation}
    \sv = \text{E}\left[\left(r-\tau\right)^2\right] = \epsilon \left(1-\epsilon\right)^2 + \left(1-\epsilon\right) \epsilon^2 = \epsilon (1-\epsilon)\,. 
\end{equation}
\subsubsection*{Systematic resampling}

Systematic resampling uses the following protocol to find $r_i$, provided $\tau_i$:
\begin{equation}
    r_i=\left|\left\{U_j : \sum_{i=1}^{j-1} \tau_i \leq U_j \leq \sum_{i=1}^{j} \tau_i \right\}\right| \text{ with } U_j = (j-1) + u_1\,,  \label{equ:ri_systematicAndStratified}
\end{equation}
where $u_1 \in [0,1]$ is a random number. Intuitively, one might convince oneself that systematic resampling also chooses $r$ as $\lfloor\tau\rfloor + 1$ with probability $\epsilon$ and $\lfloor\tau\rfloor$ with probability $1-\epsilon$ and thus should have the same sampling variance as nearest-integer resampling, i.e., $\sv = \epsilon (1-\epsilon)$.

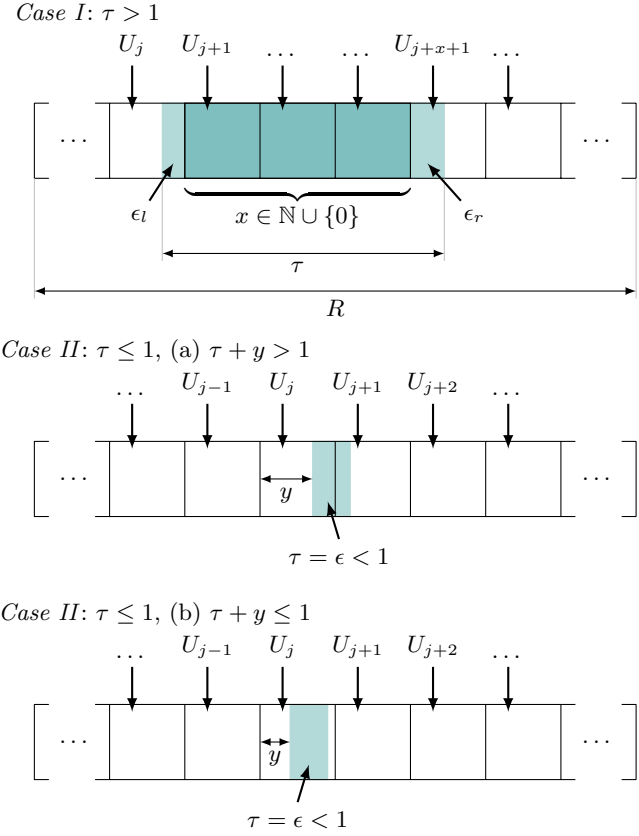
\begin{figure}
    \centering
    \begin{tikzpicture}
        \node (case1) at (-2.4,2.2) {\textit{Case I}: $\tau > 1$\hspace{1.8cm}~};
        \draw[fill=teal!30,draw=teal!30] (-2,0) rectangle (-2.3,1);
        \draw[fill=teal!30,draw=teal!30] (1.45,0) rectangle (1,1);
        \draw[fill=teal!50] (-2,0) rectangle (1,1);
        \draw[step=1] (-4,0) grid (4,1);
        \draw[latex-latex] (-2.3,-1) -- (-0.5,-1) node[below]{$\tau$} -- (1.45,-1);
        \draw[very thin,draw=lightgray] (-2.3,-1.1) -- (-2.3,0);
        \draw[very thin,draw=lightgray] (1.45,-1.1) -- (1.45,0);
        \draw[very thin,draw=lightgray] (-4,-1.6) -- (-4,0);
        \draw[very thin,draw=lightgray] (4,-1.6) -- (4,0);
        \draw[latex-latex] (-4,-1.5) -- (0,-1.5) node[below]{$R$} -- (4,-1.5);
        \draw[fill=white,draw=white] (3.2,-0.1) rectangle (3.8,1.1);
        \draw[fill=white,draw=white] (-3.8,-0.1) rectangle (-3.2,1.1);
        \node (dotsLeft) at (-3.45,0.5) {$\dots$};
        \node (dotsRight) at (3.5,0.5) {$\dots$};
        \node (epsilonLeft) at (-2.6,-0.5) {$\epsilon_l$};
        \draw[-latex,thick] (epsilonLeft) -- (-2.15,0.2);
        \node (epsilonRight) at (1.85,-0.5) {$\epsilon_r$};
        \draw[-latex,thick] (epsilonRight) -- (1.225,0.2);
        \node (x) at (-0.5,-0.4) {$\underbrace{\hspace{3cm}}_{\textstyle x\in \mathbb{N} \cup \{0\}}$};
        \draw[-latex,thick] (-2.7,1.5) node [above] {$U_j$} -- (-2.7,0.9);
        \draw[-latex,thick] (-1.7,1.5) node [above] {$U_{j+1}$} -- (-1.7,0.9);
        \draw[-latex,thick] (-0.7,1.5) node [above] {$\dots$} -- (-0.7,0.9);
        \draw[-latex,thick] (0.3,1.5) node [above] {$\dots$} -- (0.3,0.9);
        \draw[-latex,thick] (1.3,1.5) node [above] {$U_{j+x+1}$} -- (1.3,0.9);
        \draw[-latex,thick] (2.3,1.5) node [above] {$\dots$} -- (2.3,0.9);
        
        \begin{scope}[yshift=-4.5cm]
            \node (case2a) at (-2.4,2.2) {\textit{Case II}: $\tau\leq 1$, (a) $\tau+y > 1$};
            
        \draw[fill=teal!30,draw=teal!30] (-0.3,0) rectangle (0.2,1);
        \draw[step=1] (-4,0) grid (4,1);
        \draw[fill=white,draw=white] (3.2,-0.1) rectangle (3.8,1.1);
        \draw[fill=white,draw=white] (-3.8,-0.1) rectangle (-3.2,1.1);
        \node (dotsLeft) at (-3.45,0.5) {$\dots$};
        \node (dotsRight) at (3.5,0.5) {$\dots$};
        \draw[-latex,thick] (-2.7,1.5) node [above] {$\dots$} -- (-2.7,0.9);
        \draw[-latex,thick] (-1.7,1.5) node [above] {$U_{j-1}$} -- (-1.7,0.9);
        \draw[-latex,thick] (-0.7,1.5) node [above] {$U_j$} -- (-0.7,0.9);
        \draw[-latex,thick] (0.3,1.5) node [above] {$U_{j+1}$} -- (0.3,0.9);
        \draw[-latex,thick] (1.3,1.5) node [above] {$U_{j+2}$} -- (1.3,0.9);
        \draw[-latex,thick] (2.3,1.5) node [above] {$\dots$} -- (2.3,0.9);

        \draw[-latex,thick] (0.05,-0.3) node [below] {$\tau=\epsilon< 1$} -- (-0.1,0.2);
        \draw[latex-latex] (-1,0.5) -- (-0.65,0.5) node[below]{$y$} -- (-0.3,0.5);
        \end{scope}
        \begin{scope}[yshift=-8cm]
            \node (case2b) at (-2.4,2.2) {\textit{Case II}: $\tau \leq 1$, (b) $\tau+y \leq 1$};
            \draw[fill=teal!30,draw=teal!30] (-0.6,0) rectangle (-0.1,1);
            \draw[step=1] (-4,0) grid (4,1);
            \draw[fill=white,draw=white] (3.2,-0.1) rectangle (3.8,1.1);
            \draw[fill=white,draw=white] (-3.8,-0.1) rectangle (-3.2,1.1);
            \node (dotsLeft) at (-3.45,0.5) {$\dots$};
            \node (dotsRight) at (3.5,0.5) {$\dots$};
            \draw[-latex,thick] (-2.7,1.5) node [above] {$\dots$} -- (-2.7,0.9);
            \draw[-latex,thick] (-1.7,1.5) node [above] {$U_{j-1}$} -- (-1.7,0.9);
            \draw[-latex,thick] (-0.7,1.5) node [above] {$U_j$} -- (-0.7,0.9);
            \draw[-latex,thick] (0.3,1.5) node [above] {$U_{j+1}$} -- (0.3,0.9);
            \draw[-latex,thick] (1.3,1.5) node [above] {$U_{j+2}$} -- (1.3,0.9);
            \draw[-latex,thick] (2.3,1.5) node [above] {$\dots$} -- (2.3,0.9);

            \draw[-latex,thick] (-0.5,-0.3) node [below] {$\tau=\epsilon<1$} -- (-0.35,0.2);    
            \draw[latex-latex] (-1,0.5) -- (-0.8,0.5) node[below]{$y$} --  (-0.6,0.5);
        \end{scope}
    \end{tikzpicture}
    \caption{Visualization of systematic resampling}
    \label{fig:VisSystematicRes}
\end{figure}

In the following this is shown in a rather detailed calculation which then can be used analogously in stratified resampling. Figure~\ref{fig:VisSystematicRes} shows the most important quantities used in the calculation.

The integer $x$ is equal to $\lfloor\tau\rfloor$ if $\epsilon_l+\epsilon_r<1$ and $\lfloor\tau\rfloor - 1$ otherwise. In the first case $\epsilon_l+\epsilon_r=\epsilon$, in the second case $\epsilon_l+\epsilon_r=1+\epsilon$.

\textit{Case I: $\tau > 1$.}
Let $a,b \in \{0,1\}$ such that $a$ (respectively, $b$) are equal 1 iff $U_j$ (respectively, $U_{j+x+1}$) is within the highlighted region of $\epsilon_l$ (respectively, $\epsilon_r$), i.e.,

\begin{equation}
    a = \begin{cases}
        1\,, & u_1 \geq 1-\epsilon_l \\
        0\,, & \text{else}
    \end{cases} ;\quad \quad \quad
    b = \begin{cases}
        1\,, & u_1 \leq \epsilon_r \\
        0\,, & \text{else}
    \end{cases}.
    \label{equ:DefABSyst}
\end{equation}

\noindent Then the number of replicas made is $r=x+a+b$ with some probability $P_\tau(r=k)$
. Plugging in the minimal and maximal values of $x$, $a$ and $b$ gives lower and upper bounds for $r$, namely $\lfloor\tau\rfloor - 1$ and $\lfloor\tau\rfloor+2$. From Eq.~(\ref{equ:DefABSyst}) it can be concluded that
\begin{equation}
    a + b = \begin{cases}
        2\,, & 1-\epsilon_l \leq u_1 \leq e_r \\
        1\,, & (1-\epsilon_l \leq u_1 \text{ and } \epsilon_r > u_1) \\
        ~& \text{ or } (1-\epsilon_l > u_1 \text{ and } u_1 \leq \epsilon_r)\\
        0\,, & 1-\epsilon_l > u_1 > e_r
    \end{cases}\,.
\end{equation}

\noindent Using the condition that $u_1$ is uniformly distributed on the interval $[0,1]$, one obtains the probabilities for the possible outcomes of $a+b$:
\begin{equation}
    \begin{split}
        P(a+b=2) &= \max(0,\epsilon_l+\epsilon_r-1) \\
        P(a+b=1) &= \min(1-\epsilon_l,\epsilon_r) + \min(1-\epsilon_r,\epsilon_l) \\
        P(a+b=0) &= \max(0,1-\epsilon_l-\epsilon_r)
    \end{split}
\end{equation}

\noindent These can be reformulated conveniently as
\begin{equation}
    \begin{split}
        P(a+b=2) &= \begin{cases}0\,, & \epsilon_l + \epsilon_r < 1 \\  \epsilon_l+\epsilon_r-1 = \epsilon\,, & \epsilon_l + \epsilon_r \geq 1\end{cases}\,, \\
        P(a+b=1) 
        &= \begin{cases}\epsilon_l+\epsilon_r = \epsilon\,,& \epsilon_l + \epsilon_r < 1 \\  2-\epsilon_l-\epsilon_r = 1 - \epsilon \,,& \epsilon_l + \epsilon_r \geq 1\end{cases}\,,\\
        P(a+b=0) &= \begin{cases}1 - \epsilon_l - \epsilon_r = 1 - \epsilon \,,& \epsilon_l + \epsilon_r < 1 \\ 0 \,,& \epsilon_l + \epsilon_r \geq 1\end{cases}\,. \\
    \end{split}
\end{equation}

Next, using $\tau = x + \epsilon_l + \epsilon_r$, the probability distribution $P_\tau(r=k)$ can be obtained as
\begin{equation}
    \begin{split}
        P_\tau&(r>\lfloor\tau\rfloor+2) = 0\,, \\
        P_\tau&(r=\lfloor\tau\rfloor+2) \\ &= \begin{cases} P(a+b=2|\epsilon_l+\epsilon_r<1)=0 \,,& \epsilon_l+\epsilon_r<1 \\ 0 \,,& \epsilon_l+\epsilon_r \geq 1\end{cases} \\ &= 0\,,  \\
        P_\tau&(r=\lfloor\tau\rfloor+1) \\ &= \begin{cases} P(a+b=1|\epsilon_l+\epsilon_r<1) = \epsilon \,,& \epsilon_l+\epsilon_r<1 \\ P(a+b=2|\epsilon_l+\epsilon_r\geq 1) = \epsilon \,,& \epsilon_l+\epsilon_r \geq 1 \end{cases} \\ &= \epsilon\,,\\
        P_\tau&(r=\lfloor\tau\rfloor) \\ &= \begin{cases} P(a+b=0|\epsilon_l+\epsilon_r<1) = 1 - \epsilon\,,& \epsilon_l+\epsilon_r<1 \\ P(a+b=1|\epsilon_l+\epsilon_r\geq 1) = 1-\epsilon \,,& \epsilon_l+\epsilon_r \geq 1 \end{cases} \\ &= 1 - \epsilon\,,\\
        P_\tau&(r=\lfloor\tau\rfloor-1) \\&= \begin{cases} 0 \,,& \epsilon_l+\epsilon_r<1 \\ P(a+b=0|\epsilon_l+\epsilon_r\geq 1)=0 \,,& \epsilon_l+\epsilon_r \geq 1 \end{cases} \\ &= 0\,, \\
        P_\tau&(r<\lfloor\tau\rfloor-1) = 0\,,
    \end{split}
    \label{equ:SystematicCase1ProbDist}
\end{equation}
which is identical to the distribution from nearest-integer resampling.

\textit{Case II: $\tau\leq1$.}
In the special case that $\tau<1$ and that it does not overlap an integer boundary, the above calculation does not work, as $\epsilon_l$ and $\epsilon_r$ are not well defined (case II b). Even in the case that it does overlap an integer boundary (case II a), the calculation above does not apply as the case $\epsilon_l+\epsilon_r > 1$ is impossible. Instead, a new parameter $y\in[0,1]$ is introduced (see Fig.~\ref{fig:VisSystematicRes}) which, assuming uniform distribution of $y$, can be averaged over $y$ to obtain the probability distribution $P_\tau(r=k)$ for $\tau \leq 1$. Fortunately, Eq.~(\ref{equ:SystematicCase1ProbDist}) is only a function of $\epsilon$ without any assumptions on $\epsilon_l$ or $\epsilon_r$ such that it applies as long as $\epsilon_l$ and $\epsilon_r$ are defined. Thus, in case II (a) the sampling variance is equal to the one in nearest-integer resampling.

In the case of $\epsilon_l$ and $\epsilon_r$ not being well defined, meaning that the highlighted region does not overlap an integer boundary, the argument is as follows: Under the condition that $u_1$ is equally distributed over $[0,1]$ the probability of the arrow (see Fig.~\ref{fig:VisSystematicRes}) to be within the highlighted region is $\tau=\epsilon$. Hence, again the probability distribution of systematic resampling is identical to the one in nearest-integer resampling. As the sampling variance in all cases I, II (a) and II (b) is equal to the nearest-integer $\sv$, no averaging is necessary, i.e.,
\begin{equation}
    \sv(\tau) = \sv(\epsilon) = \epsilon (1-\epsilon)\,.
\end{equation}

\subsubsection*{Stratified resampling}

Stratified resampling is quite similar to systematic resampling, with the difference that instead of using one random number $u_1$ for the entire population, for each integer strata an independent pseudorandom number $u_j\in[0,1]$ is drawn. The resampling protocol is given by
\begin{equation}
    r_i=\left|\left\{U_j : \sum_{i=1}^{j-1} \tau_i \leq U_j \leq \sum_{i=1}^{j} \tau_i \right\}\right| \text{ with } U_j = (j-1) + u_j\,.  \label{equ:ri_stratified}
\end{equation}

\noindent The calculation for stratified resampling is analogous to the one for systematic resampling just that instead of using $u_1$ for left and right boundary two independent random numbers, $u_l$ and $u_r$, are used. $u_l$ (respectively, $u_r$) corresponds to $u_j$ (respectively, $u_{j+x+1}$) in Fig.~\ref{fig:VisStratifiedRes}. 

\begin{figure}
    \centering
    \begin{tikzpicture}
        \node (case1) at (-2.4,2.2) {\textit{Case I}: $\tau > 1$\hspace{1.8cm}~};
        \draw[fill=teal!30,draw=teal!30] (-2,0) rectangle (-2.3,1);
        \draw[fill=teal!30,draw=teal!30] (1.45,0) rectangle (1,1);
        \draw[fill=teal!50] (-2,0) rectangle (1,1);
        \draw[step=1] (-4,0) grid (4,1);
        \draw[latex-latex] (-2.3,-1) -- (-0.5,-1) node[below]{$\tau$} -- (1.45,-1);
        \draw[very thin,draw=lightgray] (-2.3,-1.1) -- (-2.3,0);
        \draw[very thin,draw=lightgray] (1.45,-1.1) -- (1.45,0);
        \draw[very thin,draw=lightgray] (-4,-1.6) -- (-4,0);
        \draw[very thin,draw=lightgray] (4,-1.6) -- (4,0);
        \draw[latex-latex] (-4,-1.5) -- (0,-1.5) node[below]{$R$} -- (4,-1.5);
        \draw[fill=white,draw=white] (3.2,-0.1) rectangle (3.8,1.1);
        \draw[fill=white,draw=white] (-3.8,-0.1) rectangle (-3.2,1.1);
        \node (dotsLeft) at (-3.45,0.5) {$\dots$};
        \node (dotsRight) at (3.5,0.5) {$\dots$};
        \node (epsilonLeft) at (-2.6,-0.5) {$\epsilon_l$};
        \draw[-latex,thick] (epsilonLeft) -- (-2.15,0.2);
        \node (epsilonRight) at (1.85,-0.5) {$\epsilon_r$};
        \draw[-latex,thick] (epsilonRight) -- (1.225,0.2);
        \node (x) at (-0.5,-0.4) {$\underbrace{\hspace{3cm}}_{\textstyle x\in \mathbb{N} \cup \{0\}}$};
        \draw[-latex,thick] (-2.6834230779980112,1.5) node [above] {$U_j$} -- (-2.6834230779980112,0.9);
        \draw[-latex,thick] (-1.1943832191874408,1.5) node [above] {$U_{j+1}$} -- (-1.1943832191874408,0.9);
        \draw[-latex,thick] (-0.49325942137195566,1.5) node [above] {$\dots$} -- (-0.49325942137195566,0.9);
        \draw[-latex,thick] (0.23486972160606256,1.5) node [above] {$\dots$} -- (0.23486972160606256,0.9);
        \draw[-latex,thick] (1.1391897407035051,1.5) node [above] {$U_{j+x+1}$} -- (1.1391897407035051,0.9);
        \draw[-latex,thick] (2.25752289498801273,1.5) node [above] {$\dots$} -- (2.25752289498801273,0.9);        
        
        \begin{scope}[yshift=-4.5cm]
            \node (case2a) at (-2.4,2.2) {\textit{Case II}: $\tau\leq 1$, (a) $\tau+y > 1$};
            
        \draw[fill=teal!30,draw=teal!30] (-0.3,0) rectangle (0.2,1);
        \draw[step=1] (-4,0) grid (4,1);
        \draw[fill=white,draw=white] (3.2,-0.1) rectangle (3.8,1.1);
        \draw[fill=white,draw=white] (-3.8,-0.1) rectangle (-3.2,1.1);
        \node (dotsLeft) at (-3.45,0.5) {$\dots$};
        \node (dotsRight) at (3.5,0.5) {$\dots$};
        \draw[-latex,thick] (-2.6834230779980112,1.5) node [above] {$\dots$} -- (-2.6834230779980112,0.9);
        \draw[-latex,thick] (-1.1943832191874408,1.5) node [above] {$U_{j-1}$} -- (-1.1943832191874408,0.9);
        \draw[-latex,thick] (-0.49325942137195566,1.5) node [above] {$U_j$} -- (-0.49325942137195566,0.9);
        \draw[-latex,thick] (0.23486972160606256,1.5) node [above] {$U_{j+1}$} -- (0.23486972160606256,0.9);
        \draw[-latex,thick] (1.1391897407035051,1.5) node [above] {$U_{j+2}$} -- (1.1391897407035051,0.9);
        \draw[-latex,thick] (2.25752289498801273,1.5) node [above] {$\dots$} -- (2.25752289498801273,0.9);

        \draw[-latex,thick] (0.05,-0.3) node [below] {$\tau=\epsilon< 1$} -- (-0.1,0.2);
        \draw[latex-latex] (-1,0.5) -- (-0.65,0.5) node[below]{$y$} -- (-0.3,0.5);
        \end{scope}
        \begin{scope}[yshift=-8cm]
            \node (case2b) at (-2.4,2.2) {\textit{Case II}: $\tau \leq 1$, (b) $\tau+y \leq 1$};
            \draw[fill=teal!30,draw=teal!30] (-0.6,0) rectangle (-0.1,1);
            \draw[step=1] (-4,0) grid (4,1);
            \draw[fill=white,draw=white] (3.2,-0.1) rectangle (3.8,1.1);
            \draw[fill=white,draw=white] (-3.8,-0.1) rectangle (-3.2,1.1);
            \node (dotsLeft) at (-3.45,0.5) {$\dots$};
            \node (dotsRight) at (3.5,0.5) {$\dots$};
            \draw[-latex,thick] (-2.6834230779980112,1.5) node [above] {$\dots$} -- (-2.6834230779980112,0.9);
            \draw[-latex,thick] (-1.1943832191874408,1.5) node [above] {$U_{j-1}$} -- (-1.1943832191874408,0.9);
            \draw[-latex,thick] (-0.49325942137195566,1.5) node [above] {$U_j$} -- (-0.49325942137195566,0.9);
            \draw[-latex,thick] (0.23486972160606256,1.5) node [above] {$U_{j+1}$} -- (0.23486972160606256,0.9);
            \draw[-latex,thick] (1.1391897407035051,1.5) node [above] {$U_{j+2}$} -- (1.1391897407035051,0.9);
            \draw[-latex,thick] (2.25752289498801273,1.5) node [above] {$\dots$} -- (2.25752289498801273,0.9);

            \draw[-latex,thick] (-0.5,-0.3) node [below] {$\tau=\epsilon<1$} -- (-0.35,0.2);    
            \draw[latex-latex] (-1,0.5) -- (-0.8,0.5) node[below]{$y$} --  (-0.6,0.5);
        \end{scope}

    \end{tikzpicture}
    \caption{Visualization of stratified resampling. The only difference compared to Fig.~\ref{fig:VisSystematicRes} is the different location of the arrows $U_k$.}
    \label{fig:VisStratifiedRes}
\end{figure}

\textit{Case I: $\tau > 1$.} Quantities are defined very similarly:

\begin{equation}
    \begin{array}{llll}
        \tau &= x + \epsilon_l + \epsilon_r  &r &= x + a + b\\
        a &= \begin{cases}
            1 \,,& u_l \geq 1-\epsilon_l \\
            0 \,,& u_l < 1-\epsilon_l
        \end{cases} ;\quad\quad\quad
        &b &= \begin{cases}
            1 \,,& u_r \leq \epsilon_r \\
            0 \,,& u_r > \epsilon_r
        \end{cases}    
    \end{array}
    \label{equ:DefsStrat}
\end{equation}

\noindent From this 
\begin{equation}
    a + b = \begin{cases}
        2 \,,& 1-\epsilon_l \leq u_l  \text{ and } u_r \leq e_r \\
        1 \,,& (1-\epsilon_l \leq u_l \text{ and } \epsilon_r < u_r)\\
        ~ & \text{ or } (1-\epsilon_l > u_l \text{ and } \epsilon_r \geq u_r)\\
        0 \,,& 1-\epsilon_l > u_l  \text{ and } u_r > e_r
    \end{cases}
\end{equation}
can be found and assuming uniform distributions of $u_l$ and $u_r$ on $[0,1]$ gives

\begin{equation}
    \begin{split}
        P(a+b=2) &= \epsilon_l \epsilon_r\,, \\
        P(a+b=1) &= \epsilon_l (1-\epsilon_r) + (1-\epsilon_l) \epsilon_r = \epsilon_l + \epsilon_r - 2\epsilon_l\epsilon_r\,,\\
        P(a+b=0) &= (1-\epsilon_l) (1-\epsilon_r)\,.
    \end{split}
\end{equation}

\noindent Next, using $\tau = x + \epsilon_l + \epsilon_r$, the probability distribution $P_{\tau,\epsilon_l,\epsilon_r}(r=k)$ can be obtained, i.e.,
\begin{equation}
    \begin{split}
        P_{\tau,\epsilon_l,\epsilon_r}(r>\lfloor\tau\rfloor+2) &= 0\,, \\
        P_{\tau,\epsilon_l,\epsilon_r}(r=\lfloor\tau\rfloor+2) &= \begin{cases} \epsilon_l \epsilon_r \,,& \epsilon_l+\epsilon_r<1 \\ 0 \,,& \epsilon_l+\epsilon_r \geq 1\end{cases}\,,  \\
        P_{\tau,\epsilon_l,\epsilon_r}(r=\lfloor\tau\rfloor+1) &= \begin{cases} \epsilon_l + \epsilon_r - 2\epsilon_l\epsilon_r \,,& \epsilon_l+\epsilon_r<1 \\ \epsilon_l \epsilon_r \,,& \epsilon_l+\epsilon_r \geq 1 \end{cases}\,,\\
        P_{\tau,\epsilon_l,\epsilon_r}(r=\lfloor\tau\rfloor) &= \begin{cases} (1-\epsilon_l) (1-\epsilon_r) \,,& \epsilon_l+\epsilon_r<1 \\ \epsilon_l + \epsilon_r - 2\epsilon_l\epsilon_r \,,& \epsilon_l+\epsilon_r \geq 1 \end{cases}\,,\\
        P_{\tau,\epsilon_l,\epsilon_r}(r=\lfloor\tau\rfloor-1) &= \begin{cases} 0 \,,& \epsilon_l+\epsilon_r<1 \\ (1-\epsilon_l) (1-\epsilon_r) \,,& \epsilon_l+\epsilon_r \geq 1 \end{cases}\,, \\
        P_{\tau,\epsilon_l,\epsilon_r}(r<\lfloor\tau\rfloor-1) &= 0\,,
    \end{split}
\end{equation}
which is no longer independent of $\epsilon_l$ and $\epsilon_r$. One can regain independence of $\epsilon_l$ and $\epsilon_r$ by assuming that $\epsilon_l$ (or equivalently $\epsilon_r$) is uniformly distributed on $[0,1]$ and then averaging over $\epsilon_l$, i.e.,
\begin{equation}
    P_{\tau}(r=\lfloor\tau\rfloor+k) = \int_0^1 \mathrm{d} \epsilon_l P_{\tau,\epsilon_l,\epsilon_r(\epsilon_l)}(r=\lfloor\tau\rfloor+k)
\end{equation}
with
\begin{equation}
    \epsilon_r(\epsilon_l) = \begin{cases} \epsilon - \epsilon_l \,,& \epsilon \geq \epsilon_l \\ 1 + \epsilon - \epsilon_l \,,& \epsilon < \epsilon_l \end{cases}.
\end{equation}
    
As a consequence of the averaging procedure the possibility of $\epsilon_l+\epsilon_r>1$ is integrated into the term and thus does not apply when $\epsilon_l+\epsilon_r>1$ is impossible, i.e., in case II (a). Next, this yields the probability distribution
\begin{equation}
    \begin{split}
        P_{\tau}(r>\lfloor\tau\rfloor+2) &= 0 \,,\\
        P_{\tau}(r=\lfloor\tau\rfloor+2) &= \frac{\epsilon^3}{6}\,,\\
        P_{\tau}(r=\lfloor\tau\rfloor+1) &= \frac{(1 + \epsilon)^3}{6} - \frac 2 3 \epsilon^3\,,\\
        P_{\tau}(r=\lfloor\tau\rfloor) &= \frac{\epsilon^3}{2}-\epsilon^2+\frac 2 3\,,\\
        P_{\tau}(r=\lfloor\tau\rfloor-1) &= \frac{(1-\epsilon)^3}{6}\,,\\
        P_{\tau}(r<\lfloor\tau\rfloor-1) &= 0\,.
    \end{split}
\end{equation}
Using this distribution, the sampling variance can be calculated (for case I) as
\begin{equation}
    \sv(\tau) = \sum_{k=-1}^{2} (\epsilon-k)^2 P_{\tau}(r=\lfloor\tau\rfloor+k) = \frac 1 3\,.
\end{equation}

\noindent As complicated as the probability distribution may appear, under the assumption that $\epsilon_l$ is uniformly distributed the average sampling variance is constant at $1/3$.

\textit{Case II: $\tau\leq 1$. }
In this case, depending on $y$ and $\tau$, the highlighted region might partially occupy one or two integer strata. This yields the probability distribution
\begin{equation}
    \begin{split}
        P_{\tau,y}(r>2) &= 0,  \\
        P_{\tau,y}(r=2) &= \begin{cases} 0 \,,& y+\tau < 1, \\ (1-y) (\tau + y - 1) \,,& y+\tau \geq 1,\end{cases} \\
        P_{\tau,y}(r=1) &= \begin{cases} \tau \,,& y+\tau < 1, \\ (1-y) (2 - \tau - y) \\ + y (\tau + y - 1)  \,,& y+\tau \geq 1, \end{cases}\\
        P_{\tau,y}(r=0) &= \begin{cases} 1 - \tau \,,& y+\tau < 1, \\ y (2 - \tau - y) \,,& y+\tau \geq 1.\end{cases}
    \end{split}
\end{equation}

\noindent Assuming a uniform distribution of $y$ and integrating over $y$, i.e.,
\begin{equation}
    P_{\tau}(r=k) = \int_0^1 \mathrm{d} y P_{\tau,y}(r=k),
\end{equation}
then gives
\begin{equation}
    \begin{split}
        P_{\tau}(r>2) &= 0 \,,\\
        P_{\tau}(r=2) &= \frac{\tau^3}{6} \,,\\
        P_{\tau}(r=1) &= \left(1-\frac{\tau^2}{3}\right)\tau \,,\\
        P_{\tau}(r=0) &= \frac{\tau^3}{6} - \tau + 1\,,
    \end{split}
\end{equation}
and the sampling variance follows from
\begin{equation}
    \sv(\tau) = \sum_r (r-\tau)^2 P_{\tau}(r) = \left(\frac{\tau^2}{3} - \tau + 1\right)\tau\,.
\end{equation}
Putting both cases I and II together gives

\begin{equation}
    \sv(\tau) = \begin{cases}
        \frac 1 3\,, & \tau > 1 \\
        \left(\frac{\tau^2}{3} - \tau + 1\right)\tau \,, & \tau \leq 1
    \end{cases}.
\end{equation}

\subsubsection*{Poisson and Multinomial resampling}
For both the Poisson and the multinomial distribution the sampling variance is equal to $\langle\tau\rangle=1$ independently of temperature. 

The input argument for the Poisson distribution is $\lambda$ where $\lambda$ is the mean. In the case here $\lambda$ is equal to $\tau_k$. The variance of the Poisson distribution is also $\lambda$. Hence, $\sv(\tau)$ is equal to $\tau$ which on average is one.

Similarly, for the multinomial distribution the input parameters are $n$ and $p_k$ ($\sum_k p_k = 1$). $n$ here corresponds to the population size $R$ and $p_k = \tau_k / R$. The variance of $r_k$ is given by $n p_k ( 1 - p_k) = \tau_k (1 - \tau_k / R)$ which for $R$ large enough also gives

\begin{equation}
    \sv(\tau) = \tau\,.
\end{equation}

\noindent Again, averaging over all individuals yields
\begin{equation}
    \langle(r_k - \tau_k)^2\rangle = \frac 1 n \sum_k n p_k ( 1 - p_k) = \underbrace{\sum_k p_k}_{=1} - \underbrace{\sum_k p_k^2}_{\approx 0\text{ for large }n} \hspace{-0.4cm} \approx 1\,.    
\end{equation}

\subsubsection*{Residual resampling}
\label{app:residualResampling}
In residual resampling the resampling protocol is as follows: Each replica is replicated $\lfloor \tau \rfloor$ times which temporarily reduces the population size to $R'$. In a second step the population is brought to its original size $R$ by drawing $R-R'$ times from a multinomial distribution where each replica weight is proportional to its $\epsilon = \tau - \lfloor \tau \rfloor$.

As multinomial resampling has a sampling variance equal to the number of replicas created on average and since this average is $\epsilon$, the sampling variance for residual resampling follows
\begin{equation}
    \sv(\tau) = \epsilon\,.
\end{equation}
\noindent
Clearly, this function (also sometimes referred to as the sawtooth function) has a discontinuity for every integer value $\tau$ where it jumps from one to zero. This discontinuity in $\tau$ cascades to a discontinuity $\SV(\beta)$ when averaging over all energies, as can be seen in Figs.~\ref{fig:SW_Nivar} and \ref{fig:GE_samplingVariance}.
This only becomes apparent at low temperatures when due to the discreteness of the Ising model the energy and thus also the $\tau_k$ take only few values. A jump in the average sampling variance occurs at $\beta_{i}$ when some energy $E_j$ has $\tau_{\beta_{i-1}}(E_j) > 1$ at $\beta_{i-1}$ and $\tau_{\beta_i}(E_j) < 1$ at $\beta_{i}$. This can be observed for $\beta \gtrsim 0.8$ in Figs.~\ref{fig:SW_Nivar} and \ref{fig:GE_samplingVariance}. As the number of occupied energy levels decreases with increasing inverse temperature each jump is bigger than the previous one.

In Fig.~\ref{fig:resCostForDiffResamplingMethods} the resampling cost $\Delta\rho_t / \Delta \beta$ diverges as $0.5 / \Delta \beta$ in the limit $\Delta\beta\rightarrow 0$, where $0.5$ is the sampling variance residual resampling takes for small $\Delta\beta$. This value can be understood as follows: With all $\tau$'s close to one and on average equal to one, half of the $\tau$'s are less than one and the other half larger than and equal to one. The expected sampling variance $\sv$ for $\tau \lesssim 1$ (respectively, $\tau \gtrsim 1$) is one (respectively, zero). Thus, the measured $\sv$ is the average of zero and one, namely $0.5$.


\section{Specific-heat integral}
\label{app:ProofCvIntBound}
We aim to show that 
\begin{equation}
    \int_{\beta_1}^{\beta_2} \sqrt{C_V(\beta,V)} / \beta ~ \mathrm d \beta \leq D(\beta_1,\beta_2)
\end{equation}
for some constant $D(\beta_1,\beta_2)$, independent of the system size~$V$.

Given the integrals are finite, it follows from the Cauchy-Schwarz inequality that
\begin{equation}
    \left[\int_{\beta_1}^{\beta_2} \frac{\sqrt{C_V(\beta,V)}} {\beta} ~ \mathrm d \beta\right]^2 \leq \left(\beta_2-\beta_1\right) \int_{\beta_1}^{\beta_2} \frac{C_V(\beta,V) }{\beta^2} ~ \mathrm d \beta\,.
\end{equation}
From the definition of $C_V$ and setting the Boltzmann constant equal to one, we can express $C_V$ as

\begin{equation}
C_V = - \frac {\beta^2} {V} \frac{\partial E}{\partial \beta}\,.
\end{equation}
Thus, 

\begin{equation}
    \begin{split}
    \int_{\beta_1}^{\beta_2} C_V / \beta^2 ~ \mathrm d \beta &= - \frac 1 V \int_{\beta_1}^{\beta_2} \frac{\partial E}{\partial \beta} ~ \mathrm d \beta \\
    &= - \frac 1 V \left[E\left(\beta_2\right) - E(\beta_1)\right] \\
    &\leq e_{\max} - e_{\min}\equiv D^2 / (\beta_2 - \beta_1)\,,
    \end{split}
\end{equation}
where $e_{\min}$ and $e_{\max}$ are the lowest and highest possible energy densities, respectively. This proves the original statement and $D(\beta_1,\beta_2)=\sqrt{(e_{\max} - e_{\min}) (\beta_2 - \beta_1)}$, which is independent of system size. For the ferromagnetic Ising model on a $d$-dimensional hypercubic lattice one has $e_\mathrm{min} = -2d$ and $e_\mathrm{max} = 0$, leading to  $D(\beta_1,\beta_2) =\sqrt{2d(\beta_2-\beta_1)}$, $J$ being equal to unity.

\bibliography{references}

\begin{thebibliography}{38}%
\makeatletter
\providecommand \@ifxundefined [1]{%
 \@ifx{#1\undefined}
}%
\providecommand \@ifnum [1]{%
 \ifnum #1\expandafter \@firstoftwo
 \else \expandafter \@secondoftwo
 \fi
}%
\providecommand \@ifx [1]{%
 \ifx #1\expandafter \@firstoftwo
 \else \expandafter \@secondoftwo
 \fi
}%
\providecommand \natexlab [1]{#1}%
\providecommand \enquote  [1]{``#1''}%
\providecommand \bibnamefont  [1]{#1}%
\providecommand \bibfnamefont [1]{#1}%
\providecommand \citenamefont [1]{#1}%
\providecommand \href@noop [0]{\@secondoftwo}%
\providecommand \href [0]{\begingroup \@sanitize@url \@href}%
\providecommand \@href[1]{\@@startlink{#1}\@@href}%
\providecommand \@@href[1]{\endgroup#1\@@endlink}%
\providecommand \@sanitize@url [0]{\catcode `\\12\catcode `\$12\catcode
  `\&12\catcode `\#12\catcode `\^12\catcode `\_12\catcode `\%12\relax}%
\providecommand \@@startlink[1]{}%
\providecommand \@@endlink[0]{}%
\providecommand \url  [0]{\begingroup\@sanitize@url \@url }%
\providecommand \@url [1]{\endgroup\@href {#1}{\urlprefix }}%
\providecommand \urlprefix  [0]{URL }%
\providecommand \Eprint [0]{\href }%
\providecommand \doibase [0]{https://doi.org/}%
\providecommand \selectlanguage [0]{\@gobble}%
\providecommand \bibinfo  [0]{\@secondoftwo}%
\providecommand \bibfield  [0]{\@secondoftwo}%
\providecommand \translation [1]{[#1]}%
\providecommand \BibitemOpen [0]{}%
\providecommand \bibitemStop [0]{}%
\providecommand \bibitemNoStop [0]{.\EOS\space}%
\providecommand \EOS [0]{\spacefactor3000\relax}%
\providecommand \BibitemShut  [1]{\csname bibitem#1\endcsname}%
\let\auto@bib@innerbib\@empty
\bibitem [{\citenamefont {Janke}(2008)}]{Janke2007}%
  \BibitemOpen
  \bibfield  {author} {\bibinfo {author} {\bibfnamefont {W.}~\bibnamefont
  {Janke}},\ }\bibfield  {title} {\bibinfo {title} {{M}onte {C}arlo methods in
  classical statistical physics},\ }in\ \href@noop {} {\emph {\bibinfo
  {booktitle} {Computational {M}any-{P}article {P}hysics}}},\ \bibinfo {series
  and number} {Lect. Notes Phys. {\bf 739}},\ \bibinfo {editor} {edited by\
  \bibinfo {editor} {\bibfnamefont {H.}~\bibnamefont {Fehske}}, \bibinfo
  {editor} {\bibfnamefont {R.}~\bibnamefont {Schneider}},\ and\ \bibinfo
  {editor} {\bibfnamefont {A.}~\bibnamefont {Weiße}}}\ (\bibinfo  {publisher}
  {Springer},\ \bibinfo {address} {Berlin},\ \bibinfo {year} {2008})\ pp.\
  \bibinfo {pages} {79--140}\BibitemShut {NoStop}%
\bibitem [{\citenamefont {Hukushima}\ and\ \citenamefont
  {Iba}(2003)}]{Hukushima2003}%
  \BibitemOpen
  \bibfield  {author} {\bibinfo {author} {\bibfnamefont {K.}~\bibnamefont
  {Hukushima}}\ and\ \bibinfo {author} {\bibfnamefont {Y.}~\bibnamefont
  {Iba}},\ }\bibfield  {title} {\bibinfo {title} {Population annealing and its
  application to a spin glass},\ }\href {https://doi.org/10.1063/1.1632130}
  {\bibfield  {journal} {\bibinfo  {journal} {{AIP} Conf. Proc.}\ }\textbf
  {\bibinfo {volume} {690}},\ \bibinfo {pages} {200} (\bibinfo {year}
  {2003})}\BibitemShut {NoStop}%
\bibitem [{\citenamefont {Machta}(2010)}]{Machta2010}%
  \BibitemOpen
  \bibfield  {author} {\bibinfo {author} {\bibfnamefont {J.}~\bibnamefont
  {Machta}},\ }\bibfield  {title} {\bibinfo {title} {Population annealing with
  weighted averages: A {M}onte {C}arlo method for rough free-energy
  landscapes},\ }\href {https://doi.org/10.1103/physreve.82.026704} {\bibfield
  {journal} {\bibinfo  {journal} {Phys. Rev. E}\ }\textbf {\bibinfo {volume}
  {82}},\ \bibinfo {pages} {026704} (\bibinfo {year} {2010})}\BibitemShut
  {NoStop}%
\bibitem [{\citenamefont {Amey}\ and\ \citenamefont {Machta}(2018)}]{Amey2018}%
  \BibitemOpen
  \bibfield  {author} {\bibinfo {author} {\bibfnamefont {C.}~\bibnamefont
  {Amey}}\ and\ \bibinfo {author} {\bibfnamefont {J.}~\bibnamefont {Machta}},\
  }\bibfield  {title} {\bibinfo {title} {Analysis and optimization of
  population annealing},\ }\href {https://doi.org/10.1103/physreve.97.033301}
  {\bibfield  {journal} {\bibinfo  {journal} {Phys. Rev. E}\ }\textbf {\bibinfo
  {volume} {97}},\ \bibinfo {pages} {033301} (\bibinfo {year}
  {2018})}\BibitemShut {NoStop}%
\bibitem [{\citenamefont {Christiansen}\ \emph
  {et~al.}(2019{\natexlab{a}})\citenamefont {Christiansen}, \citenamefont
  {Weigel},\ and\ \citenamefont {Janke}}]{Christiansen2019a}%
  \BibitemOpen
  \bibfield  {author} {\bibinfo {author} {\bibfnamefont {H.}~\bibnamefont
  {Christiansen}}, \bibinfo {author} {\bibfnamefont {M.}~\bibnamefont
  {Weigel}},\ and\ \bibinfo {author} {\bibfnamefont {W.}~\bibnamefont
  {Janke}},\ }\bibfield  {title} {\bibinfo {title} {Accelerating molecular
  dynamics simulations with population annealing},\ }\href
  {https://doi.org/10.1103/physrevlett.122.060602} {\bibfield  {journal}
  {\bibinfo  {journal} {Phys. Rev. Lett.}\ }\textbf {\bibinfo {volume} {122}},\
  \bibinfo {pages} {060602} (\bibinfo {year} {2019}{\natexlab{a}})}\BibitemShut
  {NoStop}%
\bibitem [{\citenamefont {Amey}\ and\ \citenamefont {Machta}(2021)}]{Amey2021}%
  \BibitemOpen
  \bibfield  {author} {\bibinfo {author} {\bibfnamefont {C.}~\bibnamefont
  {Amey}}\ and\ \bibinfo {author} {\bibfnamefont {J.}~\bibnamefont {Machta}},\
  }\bibfield  {title} {\bibinfo {title} {Measuring glass entropies with
  population annealing},\ }\href {https://arxiv.org/abs/2103.13837} {\bibfield
  {journal} {\bibinfo  {journal} {Preprint arXiv:2103.13837}\ } (\bibinfo
  {year} {2021})}\BibitemShut {NoStop}%
\bibitem [{\citenamefont {Weigel}\ \emph {et~al.}(2017)\citenamefont {Weigel},
  \citenamefont {Barash}, \citenamefont {Borovsk{\'{y}}}, \citenamefont
  {Janke},\ and\ \citenamefont {Shchur}}]{Weigel2017}%
  \BibitemOpen
  \bibfield  {author} {\bibinfo {author} {\bibfnamefont {M.}~\bibnamefont
  {Weigel}}, \bibinfo {author} {\bibfnamefont {L.~Y.}\ \bibnamefont {Barash}},
  \bibinfo {author} {\bibfnamefont {M.}~\bibnamefont {Borovsk{\'{y}}}},
  \bibinfo {author} {\bibfnamefont {W.}~\bibnamefont {Janke}},\ and\ \bibinfo
  {author} {\bibfnamefont {L.~N.}\ \bibnamefont {Shchur}},\ }\bibfield  {title}
  {\bibinfo {title} {Population annealing: Massively parallel simulations in
  statistical physics},\ }\href
  {https://doi.org/10.1088/1742-6596/921/1/012017} {\bibfield  {journal}
  {\bibinfo  {journal} {J. Phys.: Conf. Ser.}\ }\textbf {\bibinfo {volume}
  {921}},\ \bibinfo {pages} {012017} (\bibinfo {year} {2017})}\BibitemShut
  {NoStop}%
\bibitem [{\citenamefont {Shchur}\ \emph {et~al.}(2019)\citenamefont {Shchur},
  \citenamefont {Barash}, \citenamefont {Weigel},\ and\ \citenamefont
  {Janke}}]{Shchur2018}%
  \BibitemOpen
  \bibfield  {author} {\bibinfo {author} {\bibfnamefont {L.}~\bibnamefont
  {Shchur}}, \bibinfo {author} {\bibfnamefont {L.}~\bibnamefont {Barash}},
  \bibinfo {author} {\bibfnamefont {M.}~\bibnamefont {Weigel}},\ and\ \bibinfo
  {author} {\bibfnamefont {W.}~\bibnamefont {Janke}},\ }\bibfield  {title}
  {\bibinfo {title} {Population annealing and large scale simulations in
  statistical mechanics},\ }in\ \href
  {https://doi.org/10.1007/978-3-030-05807-4_30} {\emph {\bibinfo {booktitle}
  {Communications in Computer and Information Science {\bf 965}}}}\ (\bibinfo
  {publisher} {Springer Nature Switzerland},\ \bibinfo {address} {Cham},\
  \bibinfo {year} {2019})\ pp.\ \bibinfo {pages} {354--366}\BibitemShut
  {NoStop}%
\bibitem [{\citenamefont {Weigel}(2018)}]{weigel:18}%
  \BibitemOpen
  \bibfield  {author} {\bibinfo {author} {\bibfnamefont {M.}~\bibnamefont
  {Weigel}},\ }\bibfield  {title} {\bibinfo {title} {{M}onte {C}arlo methods
  for massively parallel computers},\ }in\ \href
  {https://doi.org/10.1142/9789813232105_0006} {\emph {\bibinfo {booktitle}
  {Order, Disorder and Criticality}}},\ Vol.~\bibinfo {volume} {5},\ \bibinfo
  {editor} {edited by\ \bibinfo {editor} {\bibfnamefont {Y.}~\bibnamefont
  {Holovatch}}}\ (\bibinfo  {publisher} {World Scientific},\ \bibinfo {address}
  {Singapore},\ \bibinfo {year} {2018})\ pp.\ \bibinfo {pages}
  {271--340}\BibitemShut {NoStop}%
\bibitem [{\citenamefont {Hukushima}\ and\ \citenamefont
  {Nemoto}(1996)}]{Hukushima1996}%
  \BibitemOpen
  \bibfield  {author} {\bibinfo {author} {\bibfnamefont {K.}~\bibnamefont
  {Hukushima}}\ and\ \bibinfo {author} {\bibfnamefont {K.}~\bibnamefont
  {Nemoto}},\ }\bibfield  {title} {\bibinfo {title} {Exchange {M}onte {C}arlo
  method and application to spin glass simulations},\ }\href
  {https://doi.org/10.1143/jpsj.65.1604} {\bibfield  {journal} {\bibinfo
  {journal} {J. Phys. Soc. Jpn.}\ }\textbf {\bibinfo {volume} {65}},\ \bibinfo
  {pages} {1604} (\bibinfo {year} {1996})}\BibitemShut {NoStop}%
\bibitem [{\citenamefont {Bittner}\ \emph {et~al.}(2008)\citenamefont
  {Bittner}, \citenamefont {Nu{\ss}baumer},\ and\ \citenamefont
  {Janke}}]{Bittner2008}%
  \BibitemOpen
  \bibfield  {author} {\bibinfo {author} {\bibfnamefont {E.}~\bibnamefont
  {Bittner}}, \bibinfo {author} {\bibfnamefont {A.}~\bibnamefont
  {Nu{\ss}baumer}},\ and\ \bibinfo {author} {\bibfnamefont {W.}~\bibnamefont
  {Janke}},\ }\bibfield  {title} {\bibinfo {title} {Make life simple: Unleash
  the full power of the parallel tempering algorithm},\ }\href
  {https://doi.org/10.1103/physrevlett.101.130603} {\bibfield  {journal}
  {\bibinfo  {journal} {Phys. Rev. Lett.}\ }\textbf {\bibinfo {volume} {101}},\
  \bibinfo {pages} {130603} (\bibinfo {year} {2008})}\BibitemShut {NoStop}%
\bibitem [{\citenamefont {Weigel}\ \emph {et~al.}(2021)\citenamefont {Weigel},
  \citenamefont {Barash}, \citenamefont {Shchur},\ and\ \citenamefont
  {Janke}}]{Weigel2021}%
  \BibitemOpen
  \bibfield  {author} {\bibinfo {author} {\bibfnamefont {M.}~\bibnamefont
  {Weigel}}, \bibinfo {author} {\bibfnamefont {L.}~\bibnamefont {Barash}},
  \bibinfo {author} {\bibfnamefont {L.}~\bibnamefont {Shchur}},\ and\ \bibinfo
  {author} {\bibfnamefont {W.}~\bibnamefont {Janke}},\ }\bibfield  {title}
  {\bibinfo {title} {Understanding population annealing {M}onte {C}arlo
  simulations},\ }\href {https://doi.org/10.1103/physreve.103.053301}
  {\bibfield  {journal} {\bibinfo  {journal} {Phys. Rev. E}\ }\textbf {\bibinfo
  {volume} {103}},\ \bibinfo {pages} {053301} (\bibinfo {year}
  {2021})}\BibitemShut {NoStop}%
\bibitem [{\citenamefont {Kirkpatrick}\ \emph {et~al.}(1983)\citenamefont
  {Kirkpatrick}, \citenamefont {Gelatt},\ and\ \citenamefont
  {Vecchi}}]{Kirkpatrick1983}%
  \BibitemOpen
  \bibfield  {author} {\bibinfo {author} {\bibfnamefont {S.}~\bibnamefont
  {Kirkpatrick}}, \bibinfo {author} {\bibfnamefont {C.~D.}\ \bibnamefont
  {Gelatt}},\ and\ \bibinfo {author} {\bibfnamefont {M.~P.}\ \bibnamefont
  {Vecchi}},\ }\bibfield  {title} {\bibinfo {title} {Optimization by simulated
  annealing},\ }\href {https://doi.org/10.1126/science.220.4598.671} {\bibfield
   {journal} {\bibinfo  {journal} {Science}\ }\textbf {\bibinfo {volume}
  {220}},\ \bibinfo {pages} {671} (\bibinfo {year} {1983})}\BibitemShut
  {NoStop}%
\bibitem [{\citenamefont {Gordon}\ \emph {et~al.}(1993)\citenamefont {Gordon},
  \citenamefont {Salmond},\ and\ \citenamefont {Smith}}]{Gordon1993}%
  \BibitemOpen
  \bibfield  {author} {\bibinfo {author} {\bibfnamefont {N.}~\bibnamefont
  {Gordon}}, \bibinfo {author} {\bibfnamefont {D.}~\bibnamefont {Salmond}},\
  and\ \bibinfo {author} {\bibfnamefont {A.}~\bibnamefont {Smith}},\ }\bibfield
   {title} {\bibinfo {title} {Novel approach to nonlinear/non-gaussian bayesian
  state estimation},\ }\href {https://doi.org/10.1049/ip-f-2.1993.0015}
  {\bibfield  {journal} {\bibinfo  {journal} {{IEE} Proc. F (Radar and Signal
  Process.)}\ }\textbf {\bibinfo {volume} {140}},\ \bibinfo {pages} {107}
  (\bibinfo {year} {1993})}\BibitemShut {NoStop}%
\bibitem [{\citenamefont {Barash}\ \emph {et~al.}(2017)\citenamefont {Barash},
  \citenamefont {Weigel}, \citenamefont {Borovsk{\'{y}}}, \citenamefont
  {Janke},\ and\ \citenamefont {Shchur}}]{Barash2017}%
  \BibitemOpen
  \bibfield  {author} {\bibinfo {author} {\bibfnamefont {L.~Y.}\ \bibnamefont
  {Barash}}, \bibinfo {author} {\bibfnamefont {M.}~\bibnamefont {Weigel}},
  \bibinfo {author} {\bibfnamefont {M.}~\bibnamefont {Borovsk{\'{y}}}},
  \bibinfo {author} {\bibfnamefont {W.}~\bibnamefont {Janke}},\ and\ \bibinfo
  {author} {\bibfnamefont {L.~N.}\ \bibnamefont {Shchur}},\ }\bibfield  {title}
  {\bibinfo {title} {{GPU} accelerated population annealing algorithm},\ }\href
  {https://doi.org/10.1016/j.cpc.2017.06.020} {\bibfield  {journal} {\bibinfo
  {journal} {Comput. Phys. Commun.}\ }\textbf {\bibinfo {volume} {220}},\
  \bibinfo {pages} {341} (\bibinfo {year} {2017})}\BibitemShut {NoStop}%
\bibitem [{\citenamefont {Christiansen}\ \emph
  {et~al.}(2019{\natexlab{b}})\citenamefont {Christiansen}, \citenamefont
  {Weigel},\ and\ \citenamefont {Janke}}]{Christiansen2019}%
  \BibitemOpen
  \bibfield  {author} {\bibinfo {author} {\bibfnamefont {H.}~\bibnamefont
  {Christiansen}}, \bibinfo {author} {\bibfnamefont {M.}~\bibnamefont
  {Weigel}},\ and\ \bibinfo {author} {\bibfnamefont {W.}~\bibnamefont
  {Janke}},\ }\bibfield  {title} {\bibinfo {title} {Population annealing
  molecular dynamics with adaptive temperature steps},\ }\href
  {https://doi.org/10.1088/1742-6596/1163/1/012074} {\bibfield  {journal}
  {\bibinfo  {journal} {J. Phys.: Conf. Ser.}\ }\textbf {\bibinfo {volume}
  {1163}},\ \bibinfo {pages} {012074} (\bibinfo {year}
  {2019}{\natexlab{b}})}\BibitemShut {NoStop}%
\bibitem [{\citenamefont {Doucet}\ \emph {et~al.}(2001)\citenamefont {Doucet},
  \citenamefont {Freitas},\ and\ \citenamefont {Gordon}}]{Doucet2001}%
  \BibitemOpen
  \bibinfo {editor} {\bibfnamefont {A.}~\bibnamefont {Doucet}}, \bibinfo
  {editor} {\bibfnamefont {N.}~\bibnamefont {Freitas}},\ and\ \bibinfo {editor}
  {\bibfnamefont {N.}~\bibnamefont {Gordon}},\ eds.,\ \href
  {https://doi.org/10.1007/978-1-4757-3437-9} {\emph {\bibinfo {title}
  {Sequential {M}onte {C}arlo Methods in Practice}}}\ (\bibinfo  {publisher}
  {Springer},\ \bibinfo {address} {New York},\ \bibinfo {year}
  {2001})\BibitemShut {NoStop}%
\bibitem [{Note1()}]{Note1}%
  \BibitemOpen
  \bibinfo {note} {Furthermore, instead of restricting the pool of
  configurations from which to resample to the ones obtained after all
  equilibration steps, one can extend the pool by the configurations sampled
  during the equilibration~\cite {Rose2019,Amey2021}.}\BibitemShut {Stop}%
\bibitem [{\citenamefont {{Douc}}\ and\ \citenamefont
  {{Cappe}}(2005)}]{Cappe2005}%
  \BibitemOpen
  \bibfield  {author} {\bibinfo {author} {\bibfnamefont {R.}~\bibnamefont
  {{Douc}}}\ and\ \bibinfo {author} {\bibfnamefont {O.}~\bibnamefont
  {{Cappe}}},\ }\bibfield  {title} {\bibinfo {title} {Comparison of resampling
  schemes for particle filtering},\ }in\ \href@noop {} {\emph {\bibinfo
  {booktitle} {Proceedings of the 4th International Symposium on Image and
  Signal Processing and Analysis (ISPA '05)}}}\ (\bibinfo {year} {2005})\ pp.\
  \bibinfo {pages} {64--69}\BibitemShut {NoStop}%
\bibitem [{\citenamefont {Doucet}\ and\ \citenamefont
  {Johansen}(2011)}]{doucet2009tutorial}%
  \BibitemOpen
  \bibfield  {author} {\bibinfo {author} {\bibfnamefont {A.}~\bibnamefont
  {Doucet}}\ and\ \bibinfo {author} {\bibfnamefont {A.~M.}\ \bibnamefont
  {Johansen}},\ }\bibfield  {title} {\bibinfo {title} {Tutorial on particle
  filtering and smoothing: Fifteen years later},\ }in\ \href@noop {} {\emph
  {\bibinfo {booktitle} {Handbook of Nonlinear Filtering}}},\ \bibinfo {editor}
  {edited by\ \bibinfo {editor} {\bibfnamefont {D.}~\bibnamefont {Crisan}}\
  and\ \bibinfo {editor} {\bibfnamefont {B.}~\bibnamefont {Rozovskii}}}\
  (\bibinfo  {publisher} {Oxford University Press},\ \bibinfo {address} {New
  York, NY},\ \bibinfo {year} {2011})\ pp.\ \bibinfo {pages}
  {656--704}\BibitemShut {NoStop}%
\bibitem [{\citenamefont {Machta}\ and\ \citenamefont
  {Ellis}(2011)}]{Machta2011}%
  \BibitemOpen
  \bibfield  {author} {\bibinfo {author} {\bibfnamefont {J.}~\bibnamefont
  {Machta}}\ and\ \bibinfo {author} {\bibfnamefont {R.~S.}\ \bibnamefont
  {Ellis}},\ }\bibfield  {title} {\bibinfo {title} {Monte {Carlo} methods for
  rough free energy landscapes: {Population} annealing and parallel
  tempering},\ }\href {https://doi.org/10.1007/s10955-011-0249-0} {\bibfield
  {journal} {\bibinfo  {journal} {J. Stat. Phys.}\ }\textbf {\bibinfo {volume}
  {144}},\ \bibinfo {pages} {541} (\bibinfo {year} {2011})}\BibitemShut
  {NoStop}%
\bibitem [{\citenamefont {Wang}\ \emph {et~al.}(2015)\citenamefont {Wang},
  \citenamefont {Machta},\ and\ \citenamefont {Katzgraber}}]{Wang2015}%
  \BibitemOpen
  \bibfield  {author} {\bibinfo {author} {\bibfnamefont {W.}~\bibnamefont
  {Wang}}, \bibinfo {author} {\bibfnamefont {J.}~\bibnamefont {Machta}},\ and\
  \bibinfo {author} {\bibfnamefont {H.~G.}\ \bibnamefont {Katzgraber}},\
  }\bibfield  {title} {\bibinfo {title} {Population annealing: Theory and
  application in spin glasses},\ }\href
  {https://doi.org/10.1103/physreve.92.063307} {\bibfield  {journal} {\bibinfo
  {journal} {Phys. Rev. E}\ }\textbf {\bibinfo {volume} {92}},\ \bibinfo
  {pages} {063307} (\bibinfo {year} {2015})}\BibitemShut {NoStop}%
\bibitem [{\citenamefont {Schervish}\ and\ \citenamefont
  {DeGroot}(2014)}]{DeGroot2012}%
  \BibitemOpen
  \bibfield  {author} {\bibinfo {author} {\bibfnamefont {M.}~\bibnamefont
  {Schervish}}\ and\ \bibinfo {author} {\bibfnamefont {M.~H.}\ \bibnamefont
  {DeGroot}},\ }\href@noop {} {\emph {\bibinfo {title} {Probability and
  {S}tatistics}}}\ (\bibinfo  {publisher} {Pearson Education},\ \bibinfo
  {address} {Upper Saddle River, NJ},\ \bibinfo {year} {2014})\BibitemShut
  {NoStop}%
\bibitem [{\citenamefont {Onsager}(1944)}]{Onsager1944}%
  \BibitemOpen
  \bibfield  {author} {\bibinfo {author} {\bibfnamefont {L.}~\bibnamefont
  {Onsager}},\ }\bibfield  {title} {\bibinfo {title} {Crystal statistics. {I}.
  {A} two-dimensional model with an order-disorder transition},\ }\href
  {https://doi.org/10.1103/physrev.65.117} {\bibfield  {journal} {\bibinfo
  {journal} {Phys. Rev.}\ }\textbf {\bibinfo {volume} {65}},\ \bibinfo {pages}
  {117} (\bibinfo {year} {1944})}\BibitemShut {NoStop}%
\bibitem [{\citenamefont {Kaufman}(1949)}]{Kaufmann1949}%
  \BibitemOpen
  \bibfield  {author} {\bibinfo {author} {\bibfnamefont {B.}~\bibnamefont
  {Kaufman}},\ }\bibfield  {title} {\bibinfo {title} {Crystal statistics. {II}.
  {P}artition function evaluated by spinor analysis},\ }\href
  {https://doi.org/10.1103/PhysRev.76.1232} {\bibfield  {journal} {\bibinfo
  {journal} {Phys. Rev.}\ }\textbf {\bibinfo {volume} {76}},\ \bibinfo {pages}
  {1232} (\bibinfo {year} {1949})}\BibitemShut {NoStop}%
\bibitem [{\citenamefont {Beale}(1996)}]{Beale1996}%
  \BibitemOpen
  \bibfield  {author} {\bibinfo {author} {\bibfnamefont {P.~D.}\ \bibnamefont
  {Beale}},\ }\bibfield  {title} {\bibinfo {title} {Exact distribution of
  energies in the two-dimensional {I}sing model},\ }\href
  {https://doi.org/10.1103/physrevlett.76.78} {\bibfield  {journal} {\bibinfo
  {journal} {Phys. Rev. Lett.}\ }\textbf {\bibinfo {volume} {76}},\ \bibinfo
  {pages} {78} (\bibinfo {year} {1996})}\BibitemShut {NoStop}%
\bibitem [{\citenamefont {Ebert}\ \emph {et~al.}(2022)\citenamefont {Ebert},
  \citenamefont {Gessert},\ and\ \citenamefont {Weigel}}]{Ebert2022}%
  \BibitemOpen
  \bibfield  {author} {\bibinfo {author} {\bibfnamefont {P.}~\bibnamefont
  {Ebert}}, \bibinfo {author} {\bibfnamefont {D.}~\bibnamefont {Gessert}},\
  and\ \bibinfo {author} {\bibfnamefont {M.}~\bibnamefont {Weigel}},\
  }\bibfield  {title} {\bibinfo {title} {Weighted averages in population
  annealing: {A}nalysis and general framework},\ }\href
  {https://doi.org/10.1103/PhysRevE.106.045303} {\bibfield  {journal} {\bibinfo
   {journal} {Phys. Rev. E}\ }\textbf {\bibinfo {volume} {106}},\ \bibinfo
  {pages} {045303} (\bibinfo {year} {2022})}\BibitemShut {NoStop}%
\bibitem [{Note2()}]{Note2}%
  \BibitemOpen
  \bibinfo {note} {In the implementation of the resampling methods with
  constant population size some parts of the resampling step are performed
  sequentially.}\BibitemShut {Stop}%
\bibitem [{\citenamefont {Li}\ \emph {et~al.}(2015)\citenamefont {Li},
  \citenamefont {Villarrubia}, \citenamefont {Sun}, \citenamefont {Corchado},\
  and\ \citenamefont {Bajo}}]{Li2015a}%
  \BibitemOpen
  \bibfield  {author} {\bibinfo {author} {\bibfnamefont {T.}~\bibnamefont
  {Li}}, \bibinfo {author} {\bibfnamefont {G.}~\bibnamefont {Villarrubia}},
  \bibinfo {author} {\bibfnamefont {S.}~\bibnamefont {Sun}}, \bibinfo {author}
  {\bibfnamefont {J.~M.}\ \bibnamefont {Corchado}},\ and\ \bibinfo {author}
  {\bibfnamefont {J.}~\bibnamefont {Bajo}},\ }\bibfield  {title} {\bibinfo
  {title} {Resampling methods for particle filtering: {I}dentical distribution,
  a new method, and comparable study},\ }\href
  {https://doi.org/10.1631/fitee.1500199} {\bibfield  {journal} {\bibinfo
  {journal} {Frontiers of Information Technology {\&} Electronic Engineering}\
  }\textbf {\bibinfo {volume} {16}},\ \bibinfo {pages} {969} (\bibinfo {year}
  {2015})}\BibitemShut {NoStop}%
\bibitem [{\citenamefont {Efron}(1982)}]{Efron1982}%
  \BibitemOpen
  \bibfield  {author} {\bibinfo {author} {\bibfnamefont {B.}~\bibnamefont
  {Efron}},\ }\href@noop {} {\emph {\bibinfo {title} {The {J}ackknife, the
  {B}ootstrap, and other {R}esampling {P}lans}}}\ (\bibinfo  {publisher}
  {Society for Industrial and Applied Mathematics},\ \bibinfo {address}
  {Philadelphia, PA},\ \bibinfo {year} {1982})\BibitemShut {NoStop}%
\bibitem [{\citenamefont {Gessert}\ \emph {et~al.}(2022)\citenamefont
  {Gessert}, \citenamefont {Weigel},\ and\ \citenamefont
  {Janke}}]{Gessert2021}%
  \BibitemOpen
  \bibfield  {author} {\bibinfo {author} {\bibfnamefont {D.}~\bibnamefont
  {Gessert}}, \bibinfo {author} {\bibfnamefont {M.}~\bibnamefont {Weigel}},\
  and\ \bibinfo {author} {\bibfnamefont {W.}~\bibnamefont {Janke}},\ }\bibfield
   {title} {\bibinfo {title} {{Resampling schemes in population
  annealing{\textemdash}Numerical results}},\ }\href
  {https://doi.org/10.1088/1742-6596/2207/1/012012} {\bibfield  {journal}
  {\bibinfo  {journal} {J. Phys.: Conf. Ser.}\ }\textbf {\bibinfo {volume}
  {2207}},\ \bibinfo {pages} {012012} (\bibinfo {year} {2022})}\BibitemShut
  {NoStop}%
\bibitem [{\citenamefont {Propp}\ and\ \citenamefont
  {Wilson}(1996)}]{ProppWilsonCFTP}%
  \BibitemOpen
  \bibfield  {author} {\bibinfo {author} {\bibfnamefont {J.~G.}\ \bibnamefont
  {Propp}}\ and\ \bibinfo {author} {\bibfnamefont {D.~B.}\ \bibnamefont
  {Wilson}},\ }\bibfield  {title} {\bibinfo {title} {Exact sampling with
  coupled {M}arkov chains and applications to statistical mechanics},\ }\href
  {https://doi.org/https://doi.org/10.1002/(SICI)1098-2418(199608/09)9:1/2<223::AID-RSA14>3.0.CO;2-O}
  {\bibfield  {journal} {\bibinfo  {journal} {Random Struct. Algorithms}\
  }\textbf {\bibinfo {volume} {9}},\ \bibinfo {pages} {223} (\bibinfo {year}
  {1996})}\BibitemShut {NoStop}%
\bibitem [{Note3()}]{Note3}%
  \BibitemOpen
  \bibinfo {note} {Note that this limit is different from the one in which Eq.
  (42) of Ref.~\cite {Amey2018} was derived.}\BibitemShut {Stop}%
\bibitem [{Note4()}]{Note4}%
  \BibitemOpen
  \bibinfo {note} {Quasi-exact refers to a large but finite sample obtained by
  exact sampling.}\BibitemShut {Stop}%
\bibitem [{Note5()}]{Note5}%
  \BibitemOpen
  \bibinfo {note} {When the left and the right limits differ (residual
  resampling), then the mean of the two values is taken, cf.\ also
  Appendix~\ref {app:residualResampling}.}\BibitemShut {Stop}%
\bibitem [{Note6()}]{Note6}%
  \BibitemOpen
  \bibinfo {note} {Note that the computational cost attributed to smaller or
  larger temperature steps is \protect \emph {not} taken into account in the
  notion of resampling cost. This is particularly true as $\Delta \beta
  \rightarrow 0$.}\BibitemShut {Stop}%
\bibitem [{\citenamefont {Geary}(1935)}]{Geary1935}%
  \BibitemOpen
  \bibfield  {author} {\bibinfo {author} {\bibfnamefont {R.~C.}\ \bibnamefont
  {Geary}},\ }\bibfield  {title} {\bibinfo {title} {The ratio of the mean
  deviation to the standard deviation as a test of normality},\ }\href
  {https://doi.org/10.1093/biomet/27.3-4.310} {\bibfield  {journal} {\bibinfo
  {journal} {Biometrika}\ }\textbf {\bibinfo {volume} {27}},\ \bibinfo {pages}
  {310} (\bibinfo {year} {1935})}\BibitemShut {NoStop}%
\bibitem [{\citenamefont {Rose}\ and\ \citenamefont {Machta}(2019)}]{Rose2019}%
  \BibitemOpen
  \bibfield  {author} {\bibinfo {author} {\bibfnamefont {N.}~\bibnamefont
  {Rose}}\ and\ \bibinfo {author} {\bibfnamefont {J.}~\bibnamefont {Machta}},\
  }\bibfield  {title} {\bibinfo {title} {Equilibrium microcanonical annealing
  for first-order phase transitions},\ }\href
  {https://doi.org/10.1103/physreve.100.063304} {\bibfield  {journal} {\bibinfo
   {journal} {Phys. Rev. E}\ }\textbf {\bibinfo {volume} {100}},\ \bibinfo
  {pages} {063304} (\bibinfo {year} {2019})}\BibitemShut {NoStop}%
\end{thebibliography}%

\end{document}